\documentclass[a4paper,fleqn,usenatbib]{mnras}

\usepackage{mathptmx}
\usepackage{color}
\usepackage[T1]{fontenc}
\usepackage{ae,aecompl}

\usepackage{graphicx}	
\usepackage{amsmath}	
\usepackage{amssymb}	

\newcommand{\bc}{}

\title[Dusty disc dispersal]{Radiation pressure clear-out of dusty photoevaporating discs}

\author[Owen, J. E. \& Kollmeier, J. A.]{
James E. Owen$^{1}$\thanks{E-mail: james.owen@imperial.ac.uk}
and Juna A. Kollmeier$^{2}$
\\
$^{1}$Astrophysics Group, Imperial College London, Prince Consort Road, London SW7 2AZ, UK\\
$^{2}$Observatories of the Carnegie Institution of Washington, 813 Santa Barbara Street, Pasadena, CA 91101, USA
}

\pubyear{2015}

\begin{document}
\label{firstpage}
\pagerange{\pageref{firstpage}--\pageref{lastpage}}
\maketitle

\begin{abstract}
Theoretical models of protoplanetary disc dispersal predict a phase where photoevaporation has truncated the disc at several AU, creating a pressure trap which is dust-rich. Previous models predicted this phase could be long-lived ($\sim$~Myr), contrary to the observational constraints. We show that dust in the pressure trap can be removed from the disc by radiation pressure exerting a significant acceleration, and hence radial velocity, on small dust particles that reside in the surface layers of the disc. The dust in the pressure trap is not subject to radial drift so it can grow to reach sizes large enough to fragment. Hence small particles removed from the surface layers are replaced by the fragments of larger particles. This link means radiation pressure can deplete the dust at all particle sizes. Through a combination of 1D and 2D models, along with secular models that follow the disc's long-term evolution, we show that radiation pressure can deplete dust from pressure traps created by photoevaporation in $\sim 10^5$~years, while the photoevaporation created cavity still resides at 10s of AU. After this phase of radiation pressure removal of dust, the disc is gas-rich and dust depleted and radially optically thin to stellar light, having observational signatures similar to a gas-rich, young debris disc. Indeed many of the young stars ($\lesssim10$~Myr old) classified as hosting a debris disc may rather be discs that have undergone this process.  
\end{abstract}




\section{Introduction} \label{sec:intro}
Protoplanetary discs are the environments in which planets form and migrate. These discs contain both gas and dust particles and are observed to live for up to $\sim 10$~Myr, before being dispersed \citep[e.g.][]{Haisch2001,Hernandez2007}. {\bc Since many observed exoplanets (including the ubiquitous, close-in super-Earths and mini-Neptunes) contain voluminous hydrogen/helium rich envelopes \citep[e.g.][]{Wolfgang2015}, it is hypothesised they formed before the gas disc dispersed  \citep[e.g.][]{Owen2017b,Jankovic2018}. Therefore, understanding the disc dispersal timescale is crucial to our understanding of planet formation. }

Protoplanetary discs deplete their solid and gas reservoirs through a variety of processes. These discs are accretion discs, so gas and dust particles are accreted onto the central proto-star. Gas and dust can also be sequestered into forming planets. However, perhaps the most crucial process for disc dispersal is the loss of gas (and tiny dust particles) through a photoevaporative wind. A photoevaporative wind occurs because high-energy photons (UV and X-rays) heat-up the surface layers of the disc to temperatures of-order the escape temperature, causing the gas to escape in a thermal driven wind \citep[e.g.][]{Hollenbach1994,Font2004}. The mass-loss rates depend on which radiation bands dominate the heating, with rates in the range $10^{-10}-10^{-8}$~M$_\odot$~yr$^{-1}$ \citep[e.g.][]{Ercolano2017,Wang2017,Nakatani2018}; however, the general evolutionary pathway that photoevaporating discs follow is the same. At early times the accretion rate vastly exceeds the photoevaportive mass-loss rate, and the disc evolves as a standard accretion disc. Once the photoevaporation rate and accretion rate become comparable, then gas that would have accreted onto the star is removed from the disc in the wind, starving the inner regions. Ultimately, photoevaporation completely cuts off the supply of gas to the inner disc, opening a gap at the radius where the disc's X-ray and UV heated atmosphere is unbound enough to escape, which is typically around 1~AU $(M_*/M_\odot)$ \citep[e.g.][]{Alexander2006a,Gorti2009,Owen2011b}. Since the inner disc is cut-off from resupply, it drains onto the central star rapidly, on its local viscous time \citep{Clarke2001,Ruden2004}, leaving behind a large hole extending out from the star to a radius of order 1-10~AU, then remaining gas and dust rich disc that is photoevaporated to large radii. 

The photoevaporation model successfully explains the ``two-timescale'' nature of protoplanetary disc evolution, where the inner regions of protoplanetary discs appear to evolve slowly on Myr timescales, before dispersing on a much more rapid timescale \citep[e.g.][]{Kenyon1995,Ercolano2011,Koepferl2013,Ercolano2014}. Furthermore, slow-moving ($\sim 5-10$~km~s$^{-1}$) ionized winds are observed to be occurring in many nearby discs hosting young stars \citep[e.g.][]{Hartigan1995,Pascucci2009,Rigliaco2013} and are consistent with the photoevaporation model \citep{Alexander2008,Ercolano2010,Pascucci2011,Owen2013,Ercolano2016}. The photoevaporation model can also explain a large fraction of observed ``transition discs'' \citep[e.g.][]{Owen2012,Espaillat2014}, specifically those with holes $\lesssim 10$~AU and accretion rates $\lesssim 10^{-9}$~M$_\odot$~yr$^{-1}$ \citep{Owen2011b} and even those with larger holes and higher accretion rates in more recent models that incorporate CO depletion in the outer disc \citep{Ercolano2018}. Transition discs are protoplanetary discs with evidence for a large hole or cavity in their discs \citep[e.g.][]{Espaillat2014}, but they are known to be a heterogeneous class of objects \citep[e.g.][]{Owen2012,vanderMarel2016} and their origins are not always clear. However, a specific prediction of the standard photoevaporation scenario is that there should be a large number of transition discs with hole sizes $\gtrsim 10$~AU but that are no longer accreting. This final long-lived stage of disc dispersal gives rise to transition discs which have lifetimes between $10^5$ and $10^6$~years, but remain optically thick -- ``relic discs'' \citet{Owen2011b}. The long disc lifetimes emerge from the simple fact that discs store most of their mass at large radii, but photoevaporative clearing proceeds from the inside out, so it will always take longer to remove the larger disc mass that resides at larger distance. While several discs satisfy this criterion \citep{Dong2017}, the number of observed non-accreting transition discs with large holes falls far below the theoretical expectations \citep{Owen2011b,Owen2012}. Studies by \citet{Cieza2013} and \citet{Hardy2015} showed that optically thick relic discs are rare and many non-accreting stars that show evidence for a circumstellar disc are more consistent with young, radially optically thin, debris discs. 

Since the relic disc phase emerges from simple arguments, it implies that some other mechanism operates to clear these discs more rapidly than originally envisioned. \citet{Owen_th12} and \citet{TS13} suggested a dynamical instability (thermal sweeping) caused by X-ray heating of the inner edge of transition discs could lead to rapid dispersal of relic discs. However, recent numerical simulations by \citet{Haworth2016} has indicated that thermal sweeping is inefficient for the majority of observed discs. Therefore, the conundrum remains.  The photoevaporation scenario produces an overproduction of non-accreting transition discs relative to observations. Here we revisit this problem on a new tack.

Many disc dispersal models have focused on the removal of the gas component; however, it is the considerably lower mass dust component of the disc that dominates the majority of the observable tracers. Here we focus on the removal of the dust from transition discs created by photoevaporation and show that radiation pressure from the central star can deplete dust discs on timescales $\sim 10^5$~yrs. 

In Section~\ref{sec:model} we provide an overview of this mechanism.  In Section~\ref{sec:Numerical} we derive the expected dust mass-loss rates for this mechanism.  In Section~\ref{sec:secular} we incorporate the numerically derived dust-mass loss rates in our disc evolution code to evaluate the secular behavior of discs in this phase and show our main results including a comparison of the expected SEDs from our model compared to observations.

\section{Overview of the Scenario} \label{sec:model}
Our new mechanism is grounded within the photoevaporative disc clearing scenario, where a combination of viscous accretion and photoevaporative mass-loss triggers inside-out disc clearing and gap opening when the accretion rate drops below the photoevaporation rate \citep[e.g.][]{Clarke2001}. In particular, all our quantitative calculations are done within the framework of the X-ray photoevaporation model \citep[][]{Owen2010,Owen2011b,Owen_th12}. 

\begin{figure}
\centering
\includegraphics[width=\columnwidth]{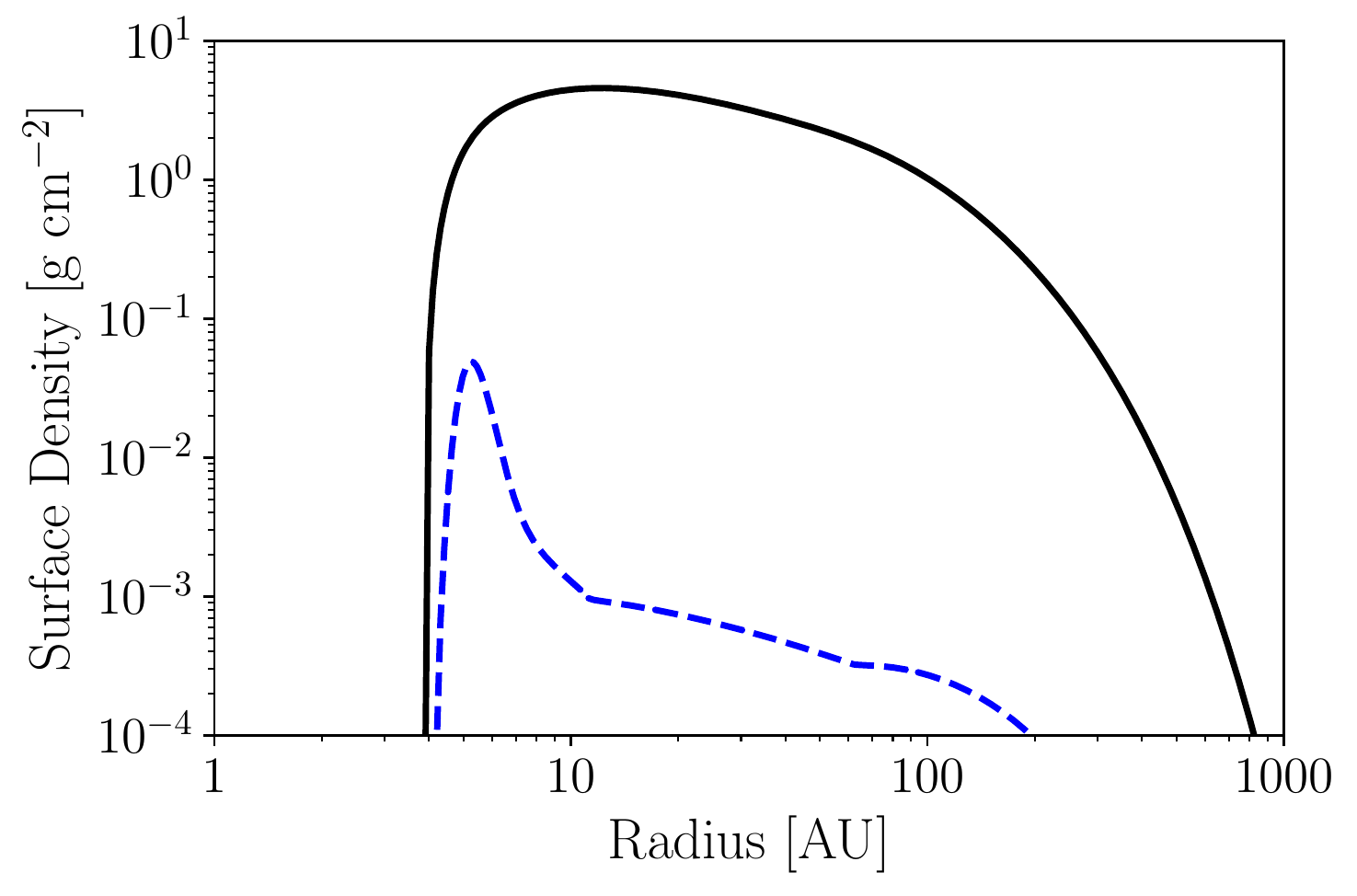}
\caption{The gas (solid) and dust (dashed) surface density distribution in a evolving and photoevaporating disc after $\sim 3.55$~Myr of evolution. The disc shown is losing mass through the X-ray photoevaporation prescription, the calculation shown is the ``median'' model described by \citet{Owen2011b}.}\label{fig:photo_model}
\end{figure}

We emphasize that our model will work in the context of any photoevaporation model (e.g. EUV/FUV), or indeed any scenario where the disc contains a significant pressure trap (see Section~\ref{sec:mmbright}).  {\bc Therefore, our scenario could be applied to models where the accretion is not driven by viscous stresses, but by magnetised winds \citep[e.g.][]{Bai2013,Bai2013b,Gressel2015,Simon2015,Bai2016,Bai2016b}. In models which attempt to describe wind-driven disc evolution, a pressure trap is typically formed after inner disc clearing \citep[e.g.][]{Wang2017b}; however, it is not clear how important our new mechanism will be without performing specific calculations\footnote{This is because our mechanism requires vertical diffusion of particles, which may be less efficient in the wind-driven accretion scenario.}.

}
After photoevaporation has opened a gap in the protoplanetary disc and the inner disc has drained onto the star, the gas surface density is maximized at a slightly larger radius than the cavity radius. This surface density maximum is important as it produces a pressure maximum in the disc. Pressure maxima are regions where there is no radial pressure support in the gas, and it rotates at the Keplerian velocity ($v_K$). Dust particles also have no pressure support, thus at the gas pressure maxima there is no drag, and dust particles tend to become trapped. In Figure~\ref{fig:photo_model}, we show the gas and dust profiles for a photoevaporating disc. This snapshot is shown after gap-opening, and the inner disc has drained onto the star. The model shown is the ``median'' model of \citet{Owen2011b} that is evolving under the X-ray photoevaporation model, see \citet{Owen2011b} and Section~\ref{sec:secular} for a detailed description of this calculation. {\bc We note that the gas column in the pressure trap is below that required for ionization from the X-rays (which has an typical absorption column of $\sim$8~g~cm$^{-2}$, e.g. \citealt{Turner2009}), implying MRI driven accretion can extend all the way to the mid-plane.}   

Figure~\ref{fig:photo_model} shows that the gas surface density profile created by photoevaporation traps a significant amount of dust in its pressure trap (located at $\sim 5.5$~AU). The pressure trap does not occur at the maximum gas surface density as there is a temperature gradient in the disc. The pressure trap has a dust surface density nearly two-orders of magnitude larger than the surrounding regions and is radially optically thick to stellar radiation. The spectral energy distributions (SEDs) of these photoevaporating discs are inconsistent with the observational constraints that show most non-accreting stars (Weak T Tauri Stars -- WTTs) either have no disc emission or show radially optically thin disc emission. Only by the time the cavity has reached radii $\sim 100$~AU, after a subsequent $\sim 1$~Myr of evolution, are these photoevaporating discs compatible with the observations \citep{Owen_th12}.   

Dust traps have another consequence: the lack of radial drift and high dust densities mean dust-particles can rapidly grow by coagulation \citep[e.g.][]{Pinilla2012}. Ultimately it is collisional induced fragmentation that limits the growth, where turbulent induced relative velocities between dust particles exceed the fragmentation velocity. This growth to the ``fragmentation barrier'' causes large particles to collide and produce smaller grains which are then able to coagulate into larger grains, so a coagulation-fragmentation equilibrium cascade is quickly established \citep{Birnstiel2011}. The maximum grain size is obtained by equating the turbulent relative velocities to the fragmentation velocity ($\sim$10-50~m~s$^{-1}$ for icy rich grains -- \citealt{Wada2008,Wada2009}) to  approximately obtain \citep[e.g.][]{Birnstiel2009}:
\begin{eqnarray}
a_{\rm frac} &=&\frac{2\Sigma_g}{\pi \alpha \rho_i}\left(\frac{u_f}{c_s}\right)^2=1\,{\rm mm} \left(\frac{\Sigma_g}{1~{\rm g~cm^{-2}}}\right)\left(\frac{\alpha}{10^{-3}}\right)^{-1}\nonumber \\ &\times&\left(\frac{\rho_i}{1.25\,{\rm g~cm^{-3}}}\right)^{-1}\left(\frac{u_f}{10~{\rm m~s^{-1}}}\right)^2\left(\frac{T}{150~{\rm K}}\right)^{-1}\label{eq:afrag}
\end{eqnarray}
where $\Sigma_g$ is the gas surface density, $\alpha$ is the Shakura-Sunyaev turbulent parameter, $\rho_i$ is the internal density of the dust particles, $u_f$ is the fragmentation velocity, $c_s$ is the sound-speed and $T$ the gas temperature. We have evaluated Equation~\ref{eq:afrag} {\bc for parameters we find in the pressure maxima of photoevaporating discs resulting from the X-ray photoevaporation model (e.g. Figure~\ref{fig:photo_model}, \citealt{Owen2011b}).

}
This growth-fragmentation cascade is key to our scenario; if small dust particles are removed through an external process, then they will be replaced by fragmentation of large particles and vice-versa. Therefore, depleting any range of particle sizes depletes {\it all} particle sizes from the pressure trap. The vertical distribution of the dust particles is size dependent. Turbulence easily lofts small dust particles, so they settle towards the mid-plane slower than large particles. Therefore the large particles (e.g. mm-sized) are typically settled towards the mid-plane, and the small particles (e.g. micron-sized) are lofted to several scale heights. For small particles with dimensionless stopping times ($\tau_s=\pi\rho_i a\Omega_K/\rho_g v_t$, with $\Omega_K$ the Keplerian angular velocity, $\rho_g$ the gas density and $v_t$ the mean thermal speed of gas particles) smaller than the viscous $\alpha$, the dust-scale height ($H_d$) relative to the gas scale height ($H$) is $H_d=H\sqrt{\alpha/\tau_s}$, \citep{Youdin2007}. Therefore, the small dust particles intercept the stellar radiation at several scale heights above the mid-plane, and it is the small dust particles that lie above the disc's photosphere.   
\subsection{The role of radiation pressure}
The dust particles that lie above the photosphere experience an additional acceleration due to radiation pressure arising from stellar photons. The strength of the radiation pressure relative to gravity can be characterized by the ratio of the radiative acceleration to the gravitational acceleration ($\beta$), where:
\begin{equation}
\beta_* = \frac{\kappa L_*}{4\pi G c M_*}
\end{equation}
$\kappa$ is opacity of an individual dust grain, $L_*$ is the stellar luminosity, $G$ is the gravitational constant, $c$ the speed of light and $M_*$ the stellar mass. In the absence of gas, particles with $\beta>0.5$ are on unbound orbits and can escape the star. However, the situation is different in the presence of gas. Small particles with $\tau_s<1$ experience gas drag that causes them to follow circular orbits; however, the reduced effective gravity (due to the radiation pressure support) they experience mean the dust-particles orbit the star slower than the gas. The gas-drag that arises causes the dust-particles to gain angular momentum from the gas and flow outwards with a velocity given by \citep{Takeuchi2003,Owen2014}:
\begin{equation}
v_d^R\approx u^R_d + \beta\tau_s v_K
\end{equation}
where $u_d^R$ is the radial gas velocity and $v_K$ is the Keplerian orbital velocity. The physics is identical to the case of gas-drag induced by pressure gradients in the gas, but in this case, it is the radiation pressure that modifies the orbital velocity of the dust, rather than the pressure gradient modifying the orbital velocity of the gas. Therefore, in the presence of gas, even particles with $\beta\ll 0.5$ can be driven away from the star. \citet{Takeuchi2003} studied the radiation pressure driven outflows in the case of optically thick primordial disc structures. They showed that in almost all cases the radiation pressure driven dust mass-flux through the optically thin surface layers was lower than the inward mass-flux of dust through the rest of the optically thick disc. 

Furthermore, in the case studied by \citet{Takeuchi2003} where the flaring photosphere means that dust particles that are driven radially outwards above the photosphere from one radius will drop below the photosphere at a slightly larger radius. In the case of a pressure trap, there will be no mass-flux through the optically thick portions of the disc. Since the pressure trap completely dominates the local dust mass (see Figure~\ref{fig:photo_model}) then the photosphere will not flare. Instead, the photosphere will follow lines of constant co-latitude (i.e. it will just be radially outwards from the star). Therefore, unlike the case in a primordial disc, dust particles that are driven radially outwards above the photosphere will always remain above the photosphere. 

The density above the photosphere is approximately $\sim 1/\kappa\ell$ (where $\ell$ is the length scale on which the density if varying). In the case of the dust trap, the length scale on which the dust density is varying is significantly smaller than the primordial disc case (which is of order $R$). The narrow dust trap results in much higher dust densities above the photosphere, and consequently, the mass-fluxes can be much larger than in the primordial disc case studied by \citet{Takeuchi2003}. 

\subsection{Disc dispersal scenario}
\begin{figure}
\centering
\includegraphics[width=\columnwidth]{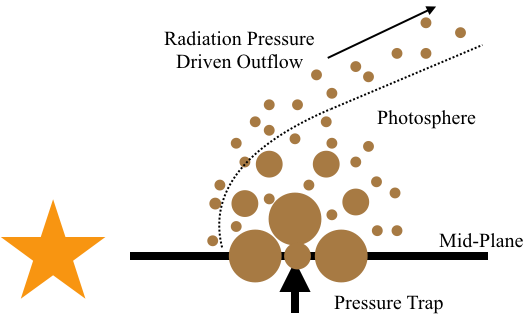}
\caption{Schematic diagram of radiation pressure driven removal of dust from a photoevaporating disc with a cavity. Large dust particles settle towards the mid-plane and the centre of the pressure trap, whereas small particles which are created by the fragmentation of the large particles are lofted above the photosphere. The small particles above the photosphere are driven outwards by radiation pressure.}\label{fig:schematic}
\end{figure}

Our new disc dispersal scenario is an extension of the previous arguments about dust trapping and the vertical distribution of particles of different sizes, but now we explicitly focus on the role of radiation pressure. Dust-particles are trapped in the pressure trap and grow to the fragmentation limit ($a_{\rm frag}$). Fragmentation creates small particles which are lofted to several scale heights by turbulence. The small particles dominate the opacity and set the photosphere's position. The small particles above the photosphere experience radiation pressure and are removed from the dust trap and disc due to the large radial velocities they attain. The removed small particles are replaced by new small particles that originate from the fragmentation of larger particles, consequently reducing the mass in the large particles. Thus, the removal of small dust-particles results in the depletion of dust particles of all sizes. This connection between the small particles and large particles only holds if the fragmentation timescale is fast enough to replenish the lost small particles. We explore this requirement further in Section~\ref{sec:discuss} in light of our results.  Our scenario is summarised in Figure~\ref{fig:schematic}. 
%
%

Radiation pressure clearing can also lead to run-away emptying of the dust-trap: as the dust-mass in the trap is depleted, the photosphere moves to smaller heights, so a more substantial fraction of the dust-mass lies above the photosphere which results in the faster clearing. Furthermore, as the photosphere moves to lower heights larger particles (which contain a more substantial fraction of the mass) also sit above the photosphere. For particles with sizes $a\gtrsim\lambda_*$ (where $\lambda_*$ is a representative wavelength of stellar photons $\sim 0.5~\mu$m), that follow the Epstein drag law, then  the dust velocity ($v_d^R\propto\beta\tau_s v_K$) is roughly independent of particle size, so the mass-loss rate accelerates as a more significant fraction of the dust mass is above the photosphere. Ultimately, the dust-trap will be depleted until the drift of dust particles from the outer portions of the disc balances the rate at which radiation pressure removes it. Figure~\ref{fig:photo_model} indicates that this will happen when the dust surface density in the trap reaches a level of order $\Sigma\sim 10^{-3}$~g~cm$^{-2}$, whence the pressure trap and the disc is radially optically thin. 

We can get a rough estimate of the dust-mass loss rate by writing it as:
\begin{equation}
\dot{M}_{\rm dust}\approx 2 \pi R_{\rm trap} H_d  \rho_{\rm phot} v_d \label{eqn:mdot_simple1}
%
%
%
%
\end{equation}
where $\rho_{\rm phot}\sim 1/\kappa H_{w}$ is the dust density above the photosphere, $H_{w}$ is the radial scale length of the dust distribution above the photosphere, $R_{\rm trap}$ is the orbital separation of the trap and $H_d$ is the dust scale height above the photosphere. Evaluating this expression for nominal parameters we find:
\begin{eqnarray}
\dot{M}_{\rm dust} &=& 3.2\times10^{-6}{\,{\rm M}_\oplus}~{\rm yr}^{-1}\,\beta\left(\frac{M_*}{1~{\rm M}_\odot}\right)^{1/2}\left(\frac{R_{\rm trap}}{5~{\rm AU}}\right)^{1/2}\nonumber\\
&\times&\left(\frac{H_d}{H_w}\right)\left(\frac{\tau_s}{10^{-2}}\right)\left(\frac{\kappa}{10^4~{\rm cm}^2~{\rm g}^{-1}}\right)^{-1}\label{eqn:mdot_simple2}
\end{eqnarray}
{\bc Note in the above, the reference opacity value is for an individual micron-sized dust grain, rather than the opacity for some gas and dust mixture.} Comparing this mass-loss rate to the mass contained in the pressure-trap in Figure~\ref{fig:photo_model} which is $\sim 0.1$~M$_\oplus$, indicates that radiation pressure is certainly capable of depleting the dust trap on a timescale $\sim 10^5$~years. However, equations~\ref{eqn:mdot_simple1}~\&~\ref{eqn:mdot_simple2} beguile the complexity of this calculation. The parameters $H_d$, $H_w$ are controlled by the radiative transfer problem, which in turn are controlled by the dust dynamics. Furthermore, the stopping time $\tau_s$ is a local stopping time (not the mid-plane value) and depends on the height of the photosphere in the disc and the opacity ($\kappa$) also depends on the particle size distribution, which also depends on the height of the photosphere. Finally, the height of the photosphere also obviously depends on the radiative transfer problem. Therefore, while the potential for rapid dust clearing due to radiation pressure clearly exists, we must ultimately appeal to numerical calculations to accurately determine the mass-fluxes. This is because several of the parameters can vary significantly, in particular, the opacity is strongly sensitive to the particle size distribution and the stopping time depends on the gas density which varies rapidly with height $(\rho_g \propto \exp(-Z^2/2H^2))$.   

\section{Radiation pressure driven outflows} \label{sec:Numerical}
The scenario we wish to numerically study is essentially shown in Figure~\ref{fig:schematic}. The goal of this section is to calculate the dust mass-loss rates so that they can be included into evolutionary calculations of viscously evolving and photoevaporating discs (Section~\ref{sec:secular}). Since the timescale to reach steady-state in the dust is likely to be much shorter than the evolutionary timescale for the gas, in all our calculations we fix the gas density profile and evolve only the dust density. We use two types of calculation, firstly 2D axis-symmetric calculations and secondly a reduced one-dimensional problem that solves for the vertical dust distribution in the dust-trap. The reason for this is that while 2D calculations are informative, we are unable to perform a large enough parameter study to determine the rate of dust removal for inclusion in long-term evolutionary calculations. Therefore, we use our 2D calculations to benchmark our basic picture of the outflows and to estimate the width of the dust-trap above the photosphere ($H_{w}$), which we need to know \emph{a priori} in the 1D models. The basic evolutionary equation for the dust is given by the following 2D advection diffusion equation \citep[e.g.][]{Takeuchi2002}:
\begin{equation}
\frac{\partial \rho_d^i}{\partial t} + \nabla \cdot \left[\rho_d^i \mathbf{v}_d^i-\frac{\rho_g\nu }{{\rm Sc}}\nabla\left(\frac{\rho_d^i}{\rho_g}\right) \right]= 0\label{eqn:2d_dust}
\end{equation}
where $\rho_d^i$ is the dust density for particles of size $a^i$, $\nu$ is the kinematic viscosity in the gas and ${\rm Sc}$ is the Schmidt number which measures the ratio of kinematic viscosity of the gas to the diffusivity of the dust. Unless otherwise explicitly stated we adopt ${\rm Sc}=1$ and work in cylindrical co-ordinates. {\bc In the above equation we have written the Schmidt number as a scalar; however, the Schmidt number need not bee the same in the vertical and radial direction. }  We follow \citet{Takeuchi2003,Owen2014} and adopt the the terminal velocity approximation, essentially assuming that the dust particles spiral inwards or outwards through circular Keplerian orbits -- an approximation that is valid when $\tau_s<1$. In this case the radial dust velocity becomes:
\begin{equation}
v^R_d=\frac{\tau_s^{-1}u^R_d+(\beta-\eta)v_K}{\tau_s+\tau_s^{-1}}\label{eqn:radial_velocity}
\end{equation}
where $\eta$ is a dimensionless measure of the pressure gradient given by:
\begin{equation}
\eta=-\frac{1}{R\Omega_K^2\rho_g}\frac{\partial P}{\partial R}
\end{equation}
Again adopting the terminal velocity approximation the vertical dust velocity becomes:
\begin{equation}
v^Z_d=-(1-\beta)\Omega_K\tau_sZ
\end{equation}
Finally, for the small particles were are interested in here the Esptein drag-law gives a dimensionless stopping time of the form:
\begin{equation}
\tau_s=\frac{\rho_ia^i\Omega_K}{\rho_gv_t}
\end{equation} 

We can reduce our 2D calculation to an approximate problem in 1D inside the dust-trap. To do this, we note that inside the dust-trap $\partial/\partial R(\rho_d^i/\rho_g)\approx0$ as the dust-density reaches a maximum inside the trap. In the optically thick regions of the disc the radial dust velocity is approximately zero inside the pressure trap. However, in the optically thin regions the radial dust velocity is approximately $v_d^R\approx\tau_s\beta v_K$. Therefore, we approximate the radial advection term as:
\begin{equation}
\frac{1}{R}\frac{\partial}{\partial R}\left(R\rho_d^iv_d^R\right)\approx \rho_d^i\tau_s\beta v_K/H_{\rm w}
\end{equation}
Therefore, the reduced 1D problem becomes:
\begin{equation}
\frac{\partial \rho_d^i}{\partial t}+\frac{\partial}{\partial Z}\left[\rho_d^iv_d^Z-\frac{\rho_g\nu}{{\rm Sc}}\frac{\partial}{\partial Z}\left(\frac{\rho_d^i}{\rho_g}\right)\right]=-\frac{\rho_d^i\tau_s\beta v_K}{H_w}\label{eqn:1d_dust}
\end{equation}
where the loss of dust due to radiation pressure now appears as a sink term in the RHS of the vertical advection-diffusion equation. With knowledge of $H_w$ then Equation~\ref{eqn:1d_dust} can be used to estimate the dust mass-loss rates from the pressure trap. We note that $H_w$ is not the width of the pressure trap in the mid-plane, but the radial scale length of the dust distribution above the photosphere and as such depends on the radiative transfer problem. We have found the value of $H_w$ cannot be estimated from the gas distribution alone and we use our full 2D calculations to calibrate an appropriate value.   
%
%

\begin{figure}
\centering
\includegraphics[width=\columnwidth]{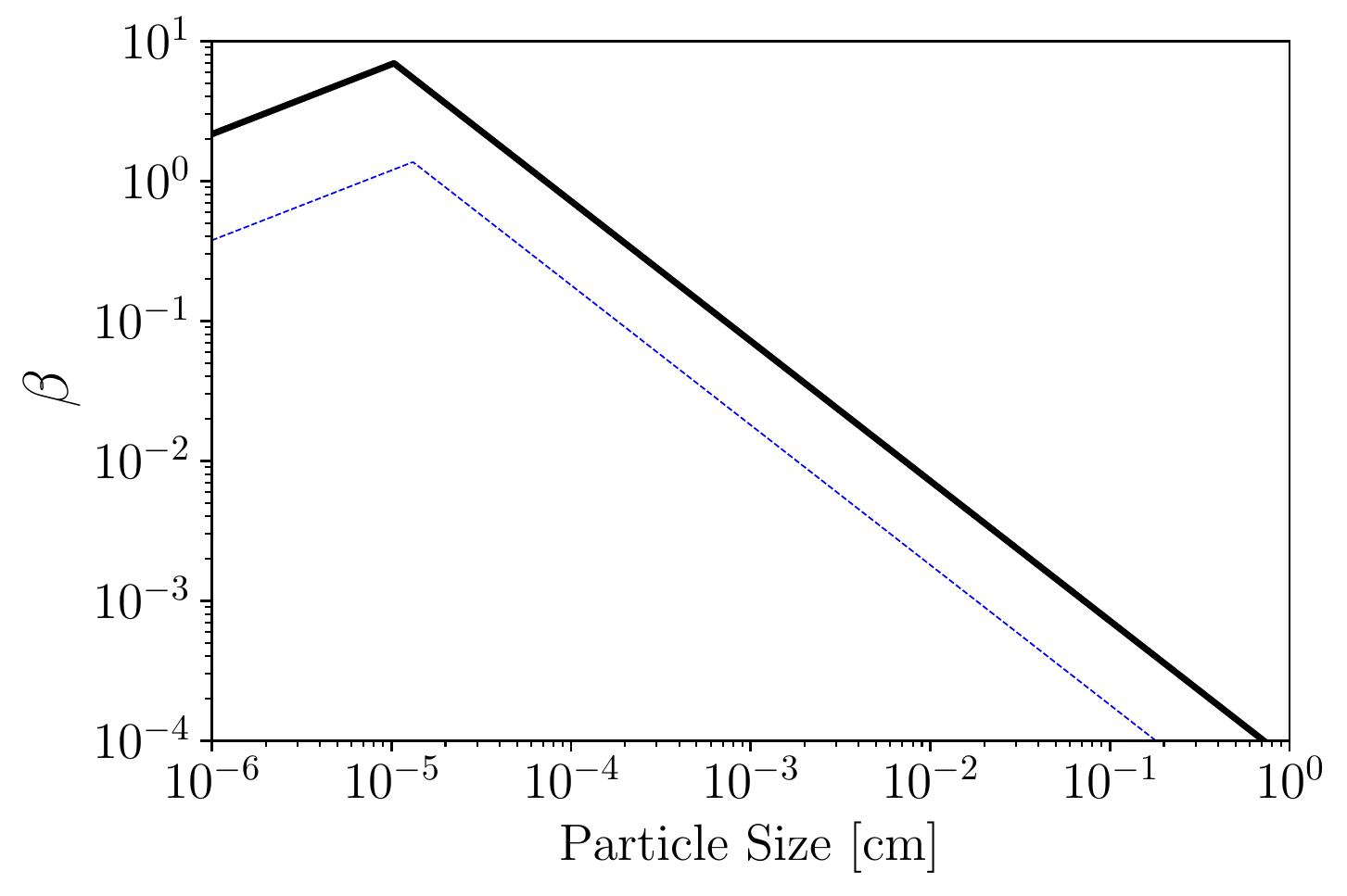}
\caption{The value of the ratio of acceleration due to radiation pressure compared to acceleration due to gravity ($\beta_*$) assuming the extinction to the star is negligible. The thick line is for a 0.7~M$_\odot$ star with a radius of 1.7~R$_\odot$ and effective temperature of 4500~K. The thin line is for a 0.5~M$_\odot$ star with a radius of 1.2~R$_\odot$ and effective temperature of 3500~K.}\label{fig:beta}
\end{figure}

In all models, we need to perform a radiative transfer calculation which requires that we know the opacity of dust particles as a function of frequency and particle size. The opacity of an individual, spherical dust grain is given by:
\begin{equation}
\kappa=\frac{3Q(a,\lambda)}{4\rho a}
\end{equation}
where $Q(a,\lambda)$ is the radiative efficiency. In our calculations we use a simplified model for the radiative efficiency of $Q=1$ for $2\pi a < \lambda$ and $Q=(2\pi a/\lambda)^{1.5}$ for $2\pi a > \lambda$, where our chosen emissivity index of 1.5 in the Rayleigh limit is close to that found for water-ice covered silicate grains \citep[e.g.][]{Chiang2001}. {\bc The use of a simplified opacity model does not account for any resonances that may exist; however, its simplicity allows us to isolate the dominate physics in our simulations.} In figure~\ref{fig:beta} we show the resulting $\beta$ parameter for a solar luminosity, $0.7$~M$_\odot$ star with an effective temperature of 4500~K, we calculate the characteristic frequency of the star using Wien's displacement law, giving $\lambda_*=0.65$~$\mu$m. The internal density of the dust particles is set to 1.25~g~cm$^{-3}$, appropriate for icy grains.
This figure indicates that small sub-micron sized dust grains are below the blow-out size; however, larger particles, e.g. 10$\mu$m sized particles still have sufficiently large $\beta$ parameters that they can reach significant radial velocities in the presence of gas. 
\subsection{Two-dimensional calculations}
We solve the two-dimensional problem on a cylindrical grid with 220 evenly spaced cells in the radial direction and 300 evenly spaced cells in the vertical direction. The parameters of the simulation setup are chosen to closely match those for the evolutionary photoevaporation calculations performed by \citet{Owen2011b}. Namely a star with mass 0.7~M$_\odot$, radius 1.7~R$_\odot$ and effective temperature of 4500~K. The gas disc has a mean-molecular-weight of 2.3 times the mass of Hydrogen and is vertically isothermal, with a radial, $R^{-1/2}$, temperature profile. The kinematic viscosity is parameterized using the standard $\alpha$ prescription with $\nu=\alpha c_s H$ and $\alpha=2.5\times10^{-3}$. {\bc This value of $\alpha$ is adopted as it is the value identified by \citet{Owen2011b} as provides agreement between the X-ray photoevaporation model and the observed protoplanetary disc population.} The gas surface density profile is taken from the viscous evolutionary calculations of \citet{Owen2011b} (i.e. the black line in Figure~\ref{fig:photo_model}) and is distributed in the vertical direction assuming hydrostatic equilibrium, in this model the pressure trap lies at approximately $5.5$~AU in the mid-plane. The inner radial boundary is set to $4\times10^{13}$~cm, and the outer boundary is at $1.75\times10^{14}$~cm, the lower vertical boundary is at the mid-plane while the higher vertical boundary is at $4.5\times10^{13}$~cm. We use reflection boundary conditions in the mid-plane and outflow boundaries on all the others.  

\begin{figure*}
\centering
\includegraphics[width=\textwidth]{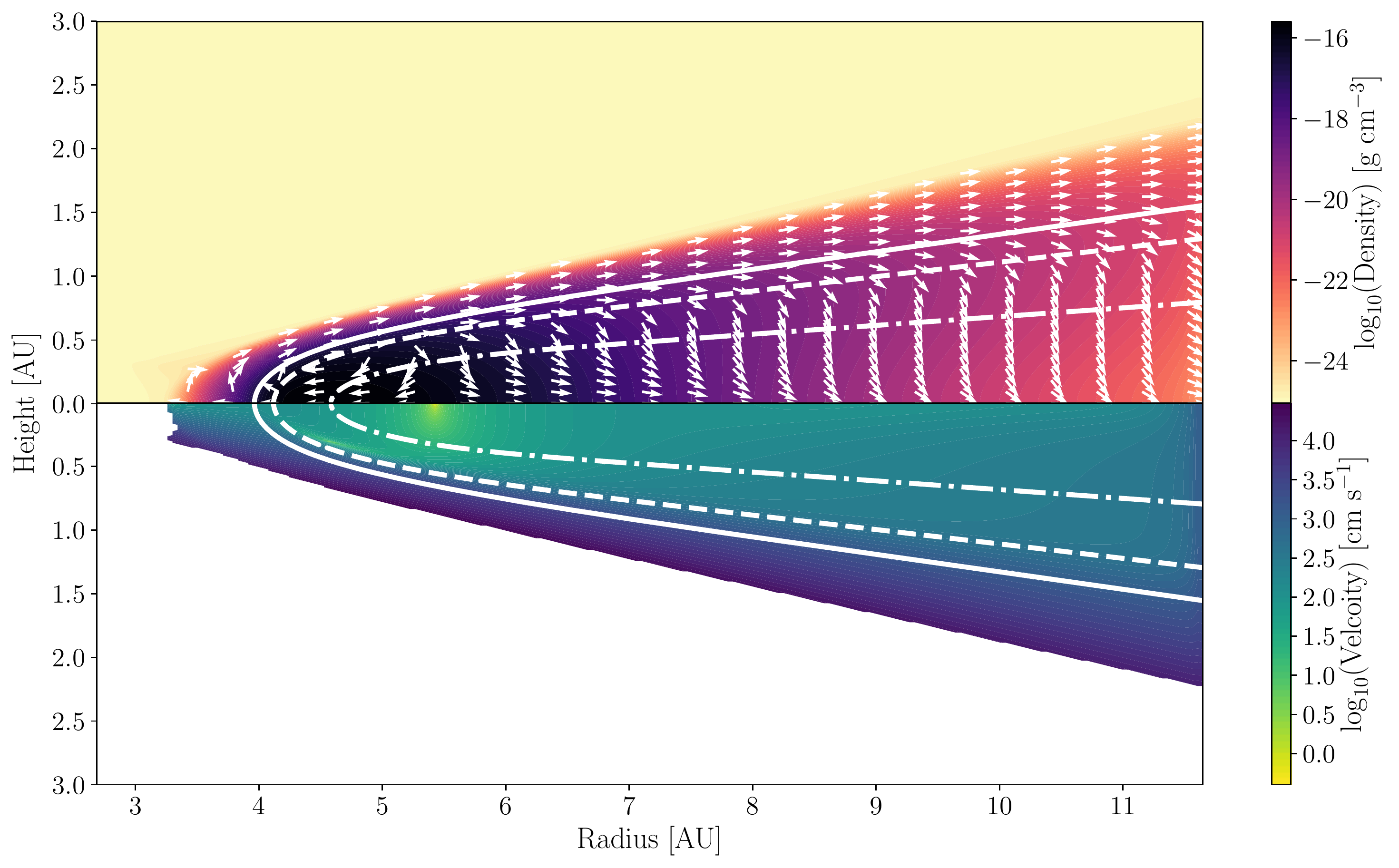}
\caption{The top panel shows the dust density and direction of the dust flux (including radiation pressure, advection, radial drift and diffusion), the bottom panel shows the magnitude of the dust velocity (in the poloidal direction) and the solid, dashed and dot-dashed lines show the positions of the $\tau_*=0.1$, 1.0 and 10 surfaces respectively. This snapshot is after $\sim 2340$ years of evolution.} \label{fig:2d_1micron}
\end{figure*}

Equation~\ref{eqn:2d_dust} is solved using an explicit integration that is first-order in time and second-order in space. The advection term is treated using a second-order upwind method that employs a van-Leer limiter. To avoid unnecessarily short-times steps that arise due to super-Keplerian radial velocities than can occur high in the disc's atmosphere we employ both a density floor of $10^{-27}$~g~cm$^{-3}$ and a maximum radial dust speed that is equal to the Keplerian-velocity. The dust density is so low at this point that these choices do not affect our results. Radiation pressure is included using a short-characteristics ray-tracing scheme where the extinction coefficient in each cell is assumed to be constant. The $\beta$ parameter is then calculated via $\beta=\beta_*\exp(-\tau_*)$ where $\tau_*$ is the optical depth to the stellar irradiation at a wavelength $\lambda_*$. The dust is initially contained within the pressure trap, such that its mid-plane density falls off as $R^{\pm 8}$ either side of the pressure trap and is vertically well mixed with the gas. The dust density in the mid-plane of the pressure trap is initialised at $10^{-12}$g~cm$^{-3}$ We integrate the problem until a quasi-steady state is reached (when the dust mass-loss rate has stabilised but is slowly evolving as dust is lost from the grid). As we do not have a dust coagulation and fragmentation routine to include in our 2d calculations, we perform these calculations with a single particle size at a time and use them to calibrate our 1D calculations. The results for 1 micron sized particles is shown in Figure~\ref{fig:2d_1micron}.
This Figure shows that our schematic picture is accurate and there is a rapid radial outflow above the photosphere. Furthermore, over a range of cavity radii, dust masses and particle sizes we find a value of $H_w\approx 0.05 R_{\rm trap}$ represents a reasonable approximation of the radial scale of the dust density above the photosphere. Therefore, in the 1D calculations detailed in Section~\ref{sec:1dmodel} we adopt this value.

\subsection{One-dimensional calculations}\label{sec:1dmodel}
\begin{figure}
\centering
\includegraphics[width=\columnwidth]{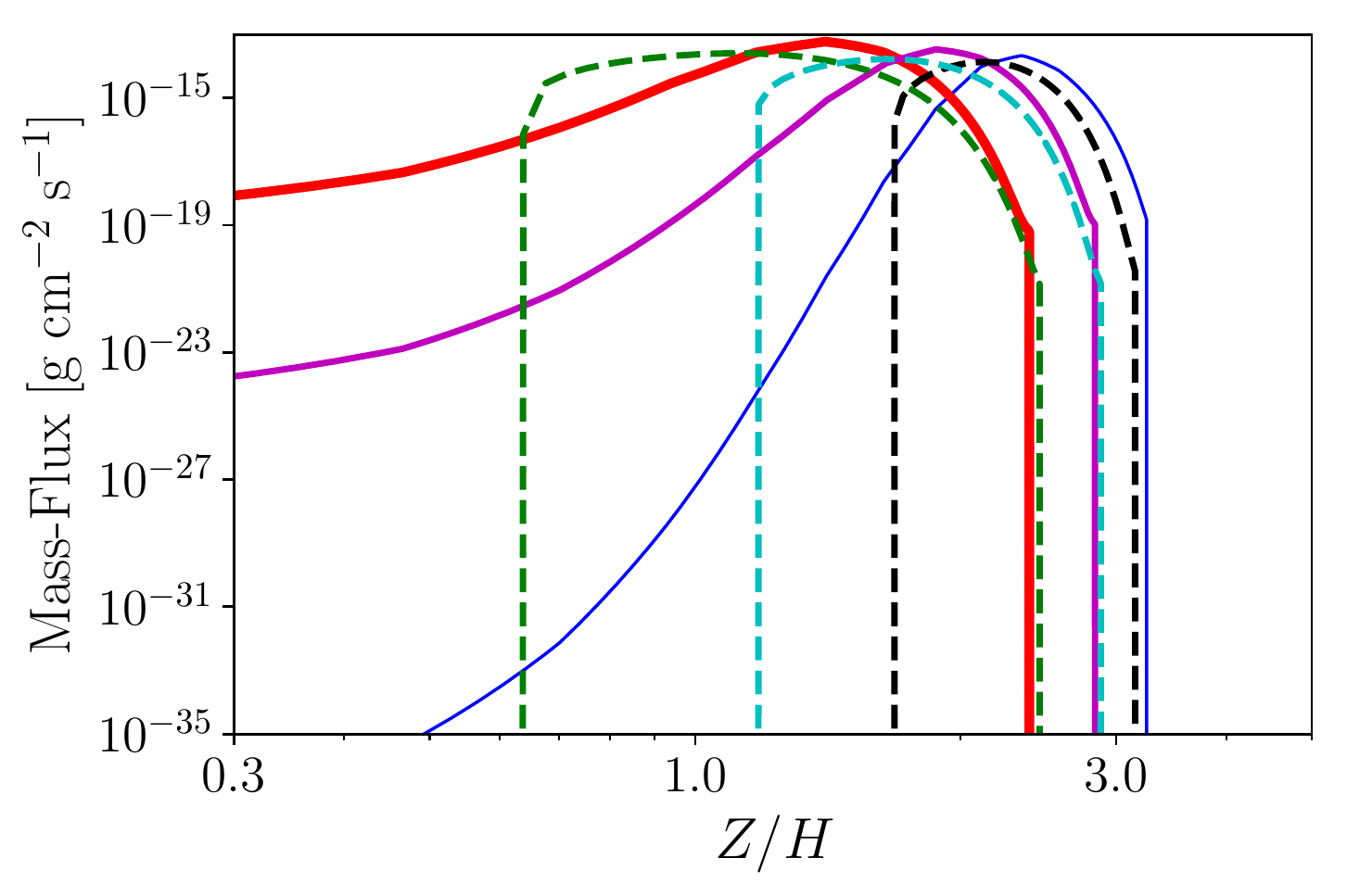}
\caption{Comparison of the mass-fluxes as a function of height taken from 1D and 2D models for particle sizes of 1 (1D - thin line, 2D - dashed black line), 3 (1D - medium thickness, 2D - dashed cyan line, and 10 microns (1D - thick line, 2D - dashed green line).}\label{fig:1d_2d_compare}
\end{figure}

To solve the reduced 1D problem (Equation~\ref{eqn:1d_dust}) we also need to know the optical depth as a function of height. To estimate this, we assume that the dust distribution has a radial distribution given by:
\begin{equation}
\rho_d^i(R,Z)=\rho_d^i(R_{\rm trap},Z)\exp\left(-\frac{(R-R_{\rm trap})^2}{2H_w^2}\right)
\end{equation}
We choose this form of the density profile as it is smooth, analytic and is of the correct form in the neighbourhood of the pressure trap. This density profile is then prescribed onto a spherical polar grid, and a ray-tracing calculation is performed to calculate the optical depth throughout this density structure. We then use cubic interpolation to calculate the values of the optical depth on the 1D vertical grid. We compare our 1D method to our 2D calculations for single particle sizes in Figure~\ref{fig:1d_2d_compare} where we show the mass-flux as a function of height above the pressure trap in both the 1D (solid lines) and 2D (dashed lines) calculations.   
We find agreement between the two simulations is good, with the general form of the mass-flux and its variation with particle size being accurately described. We find that in general, the total integrated dust mass-loss rates emanating from the dust traps agree to within $\sim 25$\% over a range of relevant particle sizes, dust-trap radii and dust densities. The rapid fall of in the dust-mass flux for the 2D case is due to inward radial drift becoming important. This feature arises from the fact that in our 2D vertically isothermal gas structures the pressure trap does not lie exactly along a line of a fixed cylindrical radius. Thus as the height is increased above the pressure trap inward radial drift results in inward transport of dust (at very low velocities) until the optical depth is low-enough for radiation pressure to dominate. 

As the first step, here we treat the coagulation-fragmentation cascade by enforcing a steady-state particle size distribution on the dust surface density. Note this does not mean we \emph{locally} enforce a particle size distribution. The vertical distribution of particles of a fixed size is free to take whatever form the balance of turbulent stirring, and gravitational settling requires. Our procedure is as follows: for each 1D simulation, we pick a minimum and maximum particle size and power-law size distribution. Then at the end of each time-step we adjust the density profiles in the following manner: we require the dust surface density distribution to follow a power-law profile in size such that $\Sigma_d(a){\rm d}a\propto m(a) a^{-p}{\rm d}a$, where $m(a)$ is the mass of a particle of size $a$. Comparing this required surface-density profile to the current surface density profile one can determine a correction factor $f(a)$ needed to enforce the current dust surface density profile to become the required one. We then update the current density profile to the required one by the following expression $\rho_d^{\rm wanted}(a,Z)=f(a)\rho_d^{\rm current}(a,Z)$. {\bc This procedure allows us to maintain the correct vertical distribution for individual particle sizes, while still adjusting the dust particle size distribution to match those resulting from steady-state coagulation-fragmentation simulations}. While this procedure is no substitute for a fully consistent particle size calculation, it does provide valuable insight into the importance of various parameters as we can vary the minimum and maximum grains sizes as well as the steepness of the power-law distribution. 

To determine the dust mass-loss rates relevant for incorporation into evolutionary calculations in Section~\ref{sec:secular} we need to know the dust mass-loss rates as a function of dust surface density, orbital radius, particle size distribution parameters, gas temperature and gas surface density. Such a complete parameter study is unfeasible even with the reduced 1D problem. However, since the evolution of the dust does not affect the long-term evolution of the gas (in the ``standard'' photoevaporation scenario described here), many of these parameters are known as a function of orbital radius from the previously computed gas evolutionary calculations taken from \citet{Owen2011b}. For example, the gas surface density in the pressure trap is known as a function of orbital radius; furthermore, if we assume that turbulent fragmentation sets the maximum particle size, then Equation~\ref{eq:afrag} allows us to calculate the maximum particle size as a function of orbital separation. The gas temperature in the pressure trap can also be computed as a function of trap location. The models of \citet{Owen2011b} adopted an $R^{-1/2}$ temperature profile arising from a passively heated flared disc structure \citep[e.g.][]{Kenyon1987} where the temperature at 1~AU is set to 100~K. Such a temperature profile is applicable for the case where the dust surface density distribution is smoothly varying with radius, such that the flaring angle of the disc's photosphere with respect to the incident stellar photons is small and also slowly varying with radius \citep[e.g.][]{Chiang1997}. This setup is not exactly the case for the situation in the pressure-trap where the rapidly varying dust density means the flaring angle with respect to the star will be much larger (as indicated by the small value of $H_w=0.05$), this will consequently mean the temperature in the pressure trap will be larger than the temperature profile adopted by \citet{Owen2011b}. Therefore, to account for the enhanced temperature in the pressure trap, we make use of the fact that the flaring angle depends on the radial scale on which the dust density at the photosphere is varying. In the case of a smoothly varying disc, the length scale on which the density at the photosphere is varying is $\sim R$; however, for our dust trap, it is $\sim H_w$. Thus as the disc temperature is proportional to the flaring angle to the one quarter power \citep[e.g.][]{Chiang2001}, then we can modify the \citet{Owen2011b} temperature profile by a factor of $(R/H_w)^{1/4}$ to find the gas temperature in the dust trap as:
\begin{equation}
T_{\rm trap}=100\,{\rm K}\left(\frac{R}{H_w}\right)^{1/4}\left(\frac{R_{\rm trap}}{1~{\rm AU}}\right)^{-1/2}
\end{equation}
This results in an approximately factor two increase in the temperature compared to the smooth primordial disc and is consistent with the enhanced MIR emission from the inner edges of transition discs \citep[e.g.][]{Ercolano2015}. 

\begin{figure}
\centering
\includegraphics[width=\columnwidth]{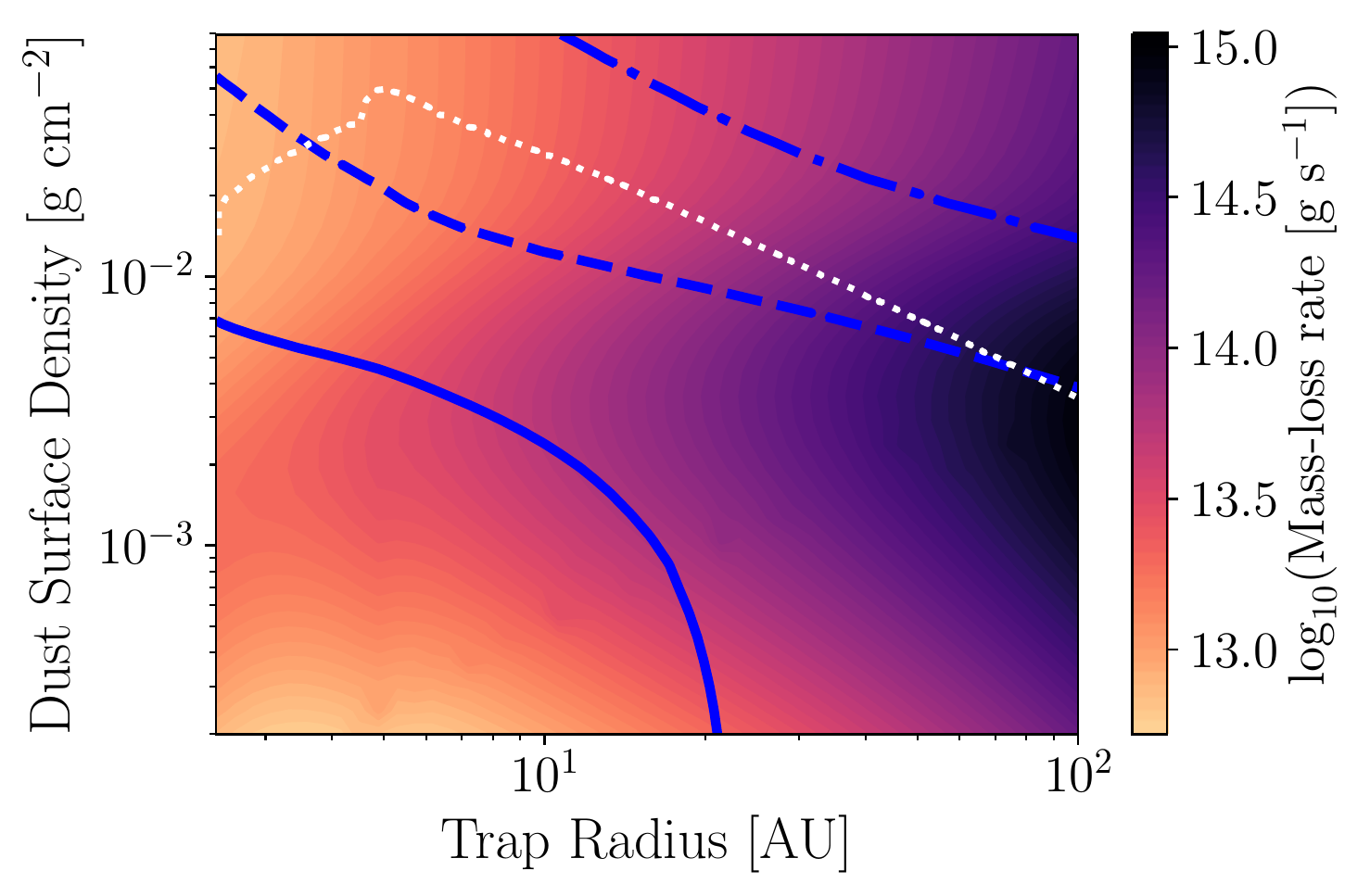}
\caption{The colour map shows the dust mass-loss rates as a function of pressure trap radius and dust surface density. The thick contours show the contours of fixed clearing time for $10^4$ (solid), $10^5$ (dashed) and $10^6$ years (dot-dashed). The dotted contour shows the trace of dust surface density as a function of trap radius for the case with no radiation pressure driven dust mass-loss. {\bc In regions where the dust surface density curve (dotted line) lies below the $10^5$~year line rapid clearing, consistent with the observations, is possible. Therefore, for this model rapid clearing is possible soon after gap opening when the trap radius is $\lesssim 5~$AU.} The trap becomes radially optically thin around a surface density of $10^{-3}$~g~cm$^{-2}$. } \label{fig:nominal}
\end{figure}

With the gas surface density, maximum particle size and gas temperature known as a function of trap location we then use our 1D models to compute the dust mass-loss rates as a function of trap location and dust surface density. In our nominal model we further adopt a minimum particle size of 0.1~$\mu$m, and an MRN dust-mass power-law index of $p=3.5$ \citep{Mathis1977}; however, we vary these parameters (and others) below. The resulting mass-loss rates for this nominal case are shown in Figure~\ref{fig:nominal}.
{\bc This Figure also includes lines showing the evolution of the dust surface-density in the trap with orbital separation and contours of constant clearing time (e.g. $M_{\rm dust}/\dot{M}$). If the dust surface density curve lies below a clearing time curve this indicates that clearing of the dust-trap on this timescale is possible\footnote{Note it is only ``possible'' as the trap could be replenished with dust by radial drift as discussed in Section~\ref{sec:secular}.}. To be consistent with the observations, the dust surface-density curve should lie below the $10^5$~year clearing timescale contour.} Figure~\ref{fig:nominal} indicates that radiation pressure driven mass loss is certainly capable of depleted photoevaporative dust trap in of order $10^5$~years, and certainly before the cavity reaches large radii, as it takes $\sim$1~Myr for the disc to be photoevaporated out to $\sim 100$~AU. The fact that the dust-surface density evolution as a function of cavity radius crosses lines of constant clear-out time multiple times indicates that the mass-loss that occurs in the early stage is important. This evolutionary pathway arises because after gap opening and before inner disc draining the photoevaporation profile is smooth (as XUV photons cannot directly impinge on the mid-plane of the outer disc). However, after the inner disc drains onto the star and direct irradiation of the outer disc's mid-plane is possible the photoevaporation profile is much more concentrated around the disc's inner edge \citep[e.g.][]{Alexander2006a,Owen2010}. This switch in photoevaporation profile sharpens the pressure gradient in the vicinity of the pressure trap, causing the evolution of the dust-surface density to peak around the point where the inner disc has finished draining onto the star. So if enough dust-mass can be removed during the earliest phase of disc clearing, then the clearing of the pressure trap may be very rapid. Alternatively, the trap could become slightly longer lived. We note that the dust mass-loss rates drop below surface densities of $\sim 10^{-3}$~g~cm$^{-2}$ as the trap is radially optically thin and the mass-loss rates then depend on the total amount of mass currently in the trap. 

\begin{figure*}
\centering
\includegraphics[width=\textwidth]{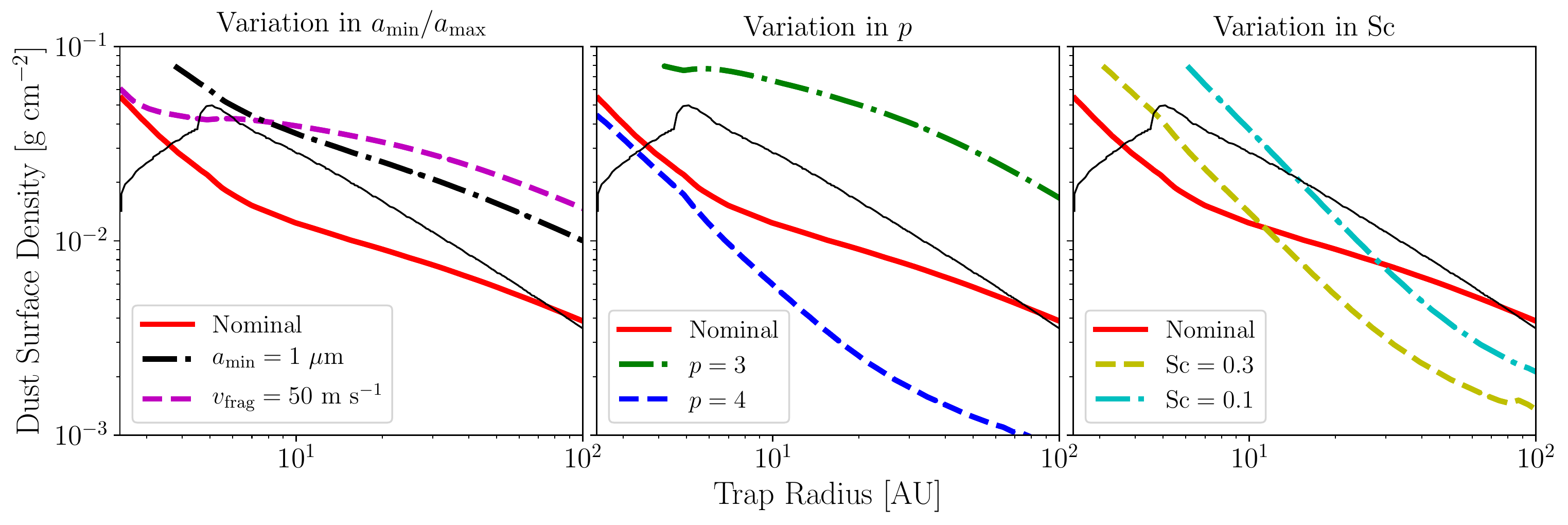}
\caption{The variation in the $10^5$~year clearing time, with auxiliary parameters. The thick solid line is the $10^5$~year clearing line from Figure~\ref{fig:nominal}.  The thin solid line is the evolution of the dust surface density in the trap without mass-loss. {\bc As in Figure~\ref{fig:nominal}, in regions where the dust surface density curve (thin solid line) lies below the $10^5$~year line rapid clearing, consistent with the observations, is possible}. The left panel considers variations in the minimum and maximum particle size: the dashed line shows a model with the fragmentation velocity increased to 50 m~s$^{-1}$ which increases the maximum particle size in the dust trap; the dash-dotted line shows a model with a minimum grain size of 1~$\mu$m. The middle panel shows variations in the power-law size index $p$ with the dot-dashed line showing a more bottom heavy distribution with $p=3$ and the dot-dashed line shows a more top-heavy distribution with $p=4$. The right-hand panel shows the effect of varying the Schmidt number with the dashed line showing ${\rm Sc}=0.3$ and the dot-dashed line showing ${\rm Sc}=0.1$.}\label{fig:vary_parameters}
\end{figure*}

We can explore the role of various other parameters in Figure~\ref{fig:vary_parameters}.
In the left panel we vary $a_{\rm min}$ and $a_{\rm max}$. The dashed line shows a model with $a_{\rm min}$ increased to 1~$\mu$m; this is a realistic possibility as Figure~\ref{fig:beta} shows that particles with sizes smaller than $\sim 1~\mu$m are below the blow-out size. Therefore, in a realistic dust coagulation-fragmentation calculation, it is possible that particles smaller than $\sim 1~\mu$m are rapidly removed before fragmentation of larger particles can replace them; this process typically sets the minimum particle size in debris discs \citep[e.g.][]{Matthews2014}. The dashed-dotted line shows a model where we increase the fragmentation speed to 50~m~s$^{-1}$, hence increasing the maximum particle size, such a fragmentation speed has been found from numerical studies of icy grains \citet{Wada2009}. Both these changes to the nominal model promote even more rapid clearing. This result arises because the opacity is dominated by the mass fraction in the smallest grains. Both these changes, increasing $a_{\rm min}$ and increasing $a_{\rm max}$, lower the opacity, hence at fixed dust surface density the total dust-mass above the photosphere and susceptible to radiation pressure is larger. In this case, both of these models would lead to very rapid clear out of the dust trap. 

The middle panel shows variations in the dust-size power-law distribution where we show values of $p=3$ (dashed), $p=3.5$ (solid, nominal model) and $p=4$ dot-dashed. As in the previous case, as one decreases the dust mass-fraction in the smallest grains (by decreasing $p$) then a larger fraction of the dust mass lies above the photosphere due to the lower opacity, leading to a more rapid clearing. 

Finally, the right panel shows the effect of changing the dust-diffusion coefficient, by decreasing the value of the Schmidt number (higher vertical diffusivity). {\bc In the outer disc non-ideal MHD effects, such as ambipolar diffusion may be important. Simulation that included ambipolar diffusion indicated enhanced diffusivity in the vertical direction \citep{Zhu2015}. Furthermore, pure hydrodynamic turbulence driven by the vertical shear instability (which can arise from a varying angular velocity with height e.g. \citealt{Nelson2013}), results in higher diffusivities in the vertical compared radial direction \citep{Stoll2017}.}  At small radii smaller Schmidt numbers result in a more rapid clearing, as the higher diffusivity negates dust settling, resulting in a higher fraction of the dust-mass above the photosphere. We summarise, that while our nominal model suggests rapid $\sim 10^5$~year clearing, most physical corrections to the chosen parameters, except for a top-heavy particle size distribution result in an even more rapid clearing.

\section{Secular Evolution}\label{sec:secular}
To asses whether radiation pressure can clear out dust from a pressure-trap created by photoevaporation, we include our dust-mass loss rates in long-term secular evolution calculations of the gas, dust and particle size distributions. We use the gas and dust evolutionary code detailed by \citet{Owen2014}, which has been modified by \citet{Jankovic2018}  to include the \citet{Birnstiel2012} particle size evolutionary model. Those previous works include detailed discussions of the numerical scheme and algorithms which we do not repeat here. Essentially we solve the evolution equation for the gas surface density:
\begin{equation}
\frac{\partial \Sigma_g}{\partial t}=\frac{3}{R}\frac{\partial}{\partial R}\left[R^{1/2}\frac{\partial}{\partial R}\left(\Sigma_g\nu R^{1/2}\right)\right]-\dot{\Sigma}_w \label{eq:s_gas}
\end{equation}
where the $\dot{\Sigma}_w$ represents the gas lost due to photoevaporation. The form of $\dot{\Sigma}_w$ is taken from the fits to numerical simulations for the X-ray photoevaporation model by \citet{Owen_th12}. The setup is identical to the ``median'' model described by \cite{Owen2011b}. The surface density is initially a zero-time Lynden-Bell \& Pringle similarity solution with initial disc mass of $0.07$~M$_\odot$ and scale radius of $18$~AU. We adopt a power-law viscosity of the form $\nu\propto R$ which is normalised such that the viscous $\alpha$ parameter is $2.5\times10^{-3}$ at 1~AU. Coupled with the gas equation, we also solve for the evolution of the dust-surface density as:
\begin{equation}
\frac{\partial \Sigma_d}{\partial t}+\frac{1}{R}\frac{\partial}{\partial R}\left[R\Sigma_d v_d^R-\nu R \Sigma_g\frac{\partial}{\partial R}\left(\frac{\Sigma_d}{\Sigma_g}\right)\right]=-\dot{\Sigma}_{\rm d} \label{eqn:s_dust}
\end{equation}
here $v_d$ is the standard vertically, and particle-sized averaged form of Equation~\ref{eqn:radial_velocity} detailed by \citet{Birnstiel2012}, note we do not include the radiation pressure term in this equation as it is focused on the optically thick mid-plane. The term $\dot{\Sigma}_{\rm d}$ represents the mass-loss from the pressure trap due to radiation pressure, we neglect dust mass loss due to particle entrained in the photoevaporative flow as it is negligible \citep{Owen2011a} and start with an initial local dust-to-gas ratio of 0.01 everywhere. To convert the total dust mass-loss rates obtained in the previous Section into a surface mass-loss profile, we adopt a Gaussian radial profile for the dust-mass loss centred on the pressure trap with a scale size of $H_w$. We ensure that the integrated mass-loss profile matches the required dust-mass loss rate for the location of the pressure trap and the current dust-surface density in the pressure trap. The mass-loss of dust due to radiation pressure only begins once photoevaporation has opened a gap and created a pressure-trap. 

The \citet{Birnstiel2012} model evolves the dust size distribution using a simple two population model, firstly a monomer size which we set to 0.1~$\mu$m and secondly a representative maximum grain size. The maximum grain size is set by whichever of drift limited growth, turbulent fragmentation or fragmentation arising from radial drift gives the smallest particle size. Furthermore, an initial and transitory growth phase is included as suggested by \citet{Birnstiel2012} where particles grow from the monomer size up-to one of dust size limits. 

Equations~\ref{eq:s_gas} and \ref{eqn:s_dust} are integrated on a grid that is uniformly spaced in $R^{1/3}$. The grid has an inner boundary at $3.75\times10^{11}$~cm and outer boundary at $3.75\times10^{16}$~cm and contains 1000 cells. We adopt a zero-torque boundary condition on the gas at both the inner and outer boundaries and outflow boundary conditions on the dust. 

\subsection{Results}
\begin{figure}
\centering
\includegraphics[width=\columnwidth]{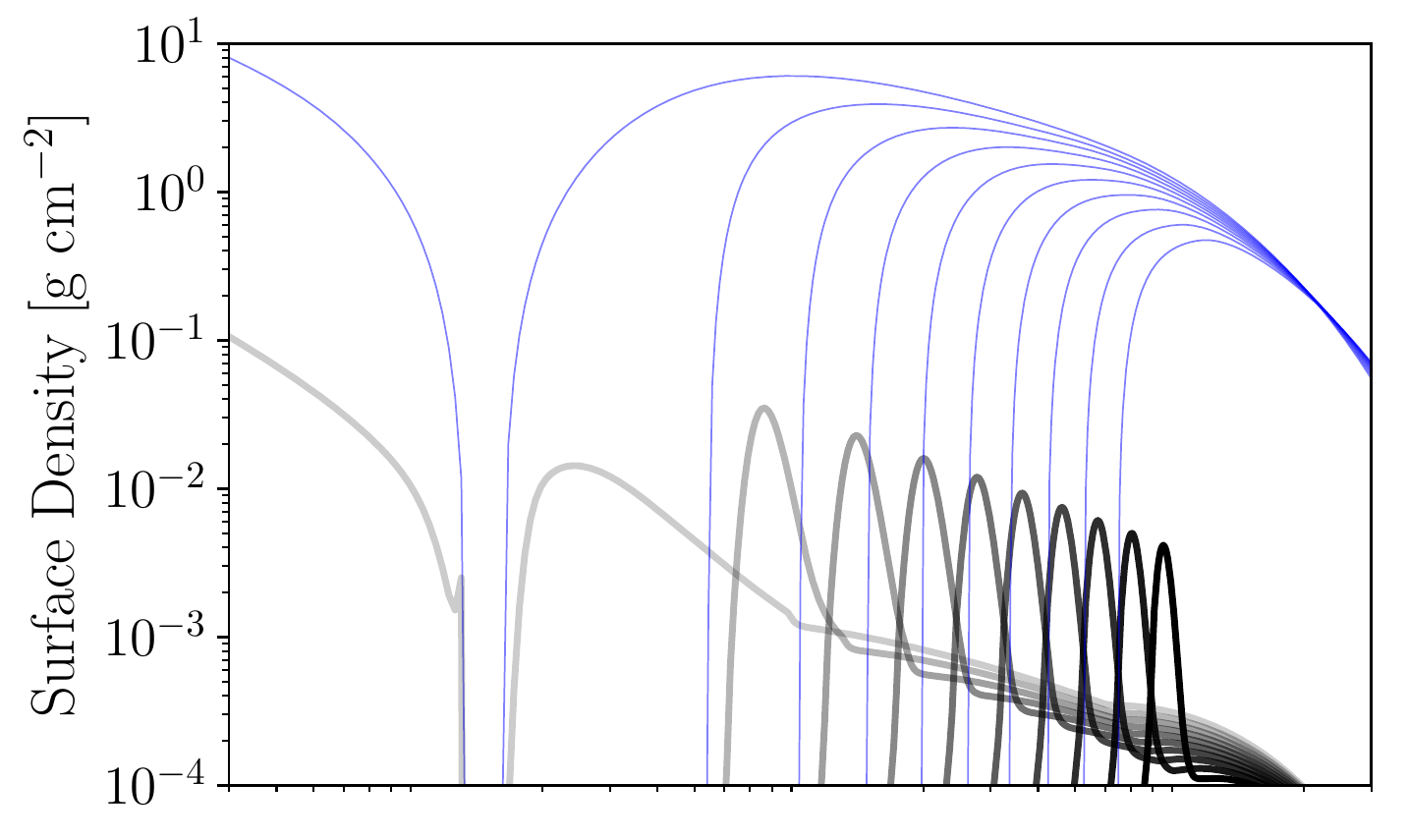}
\includegraphics[width=\columnwidth]{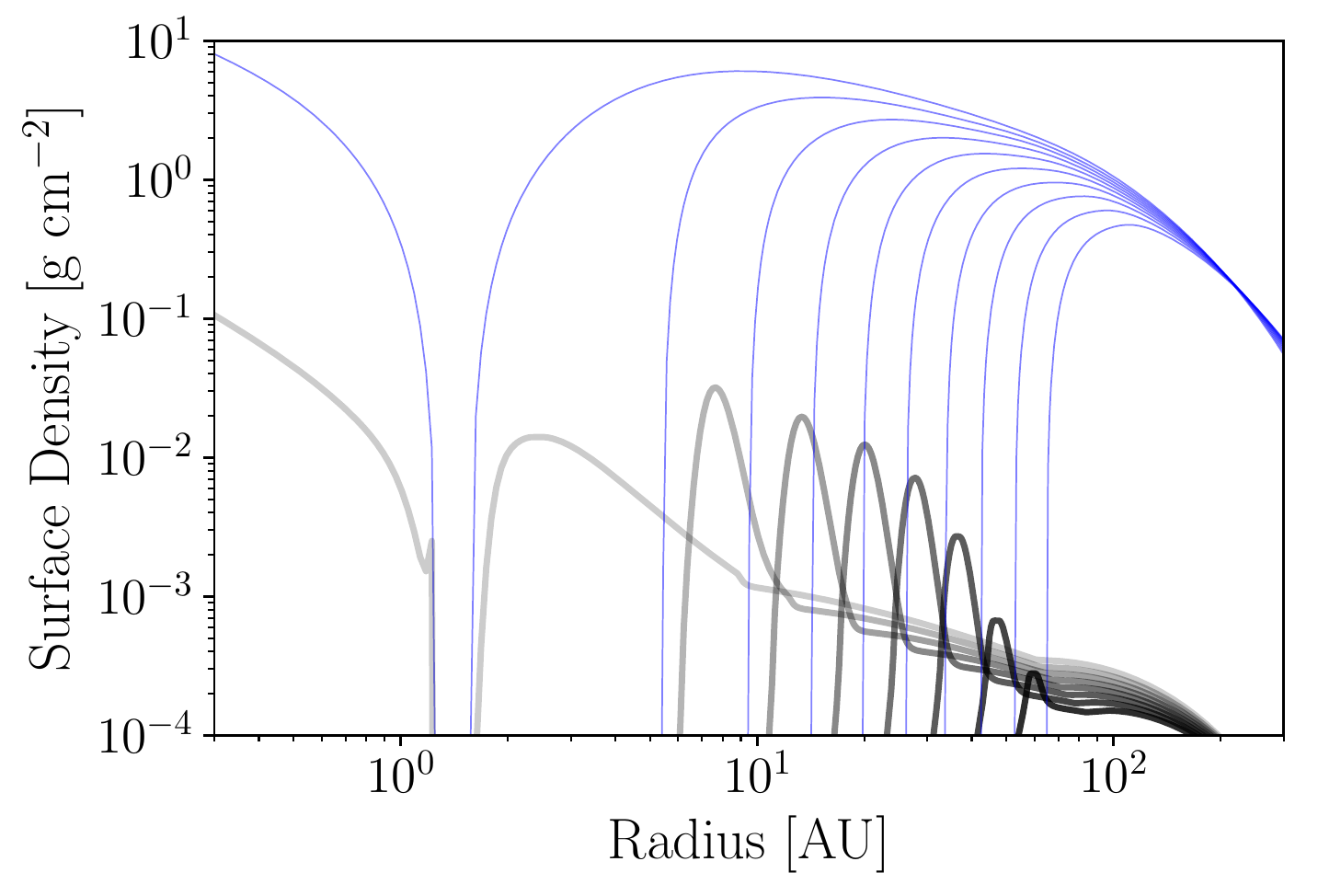}
\caption{The evolution of the dust (black, thick) and gas (blue, thin) surface densities in a photoevaporating disc. The top panel shows a model with no radiation pressure clear-out and is the previous generation of X-ray photoevaporation model. The bottom panel is the same model, but including dust mass-loss for the ``nominal'' case. The model is initially shown just after gap opening (after $\sim 3.4$~Myr of evolution) and then every 0.15Myr, gas accretion onto the star ceases after $\sim 3.55$~Myr. {\bc The darkness of the dust lines increases with time.} }\label{fig:s_evolve}
\end{figure}

In Figure~\ref{fig:s_evolve} we show the evolution of the gas and dust surface densities for a model which includes no mass-loss due to radiation pressure (top panel) compared to our nominal model (bottom panel). This Figure shows, as expected from previous considerations \citep[e.g.][]{Alexander2007,Owen_th12}, that without an extra dust mass-loss prescription photoevaporation generates dust-traps that are optically thick to a large radius (the dust trap is optically thick is it has a surface density $\gtrsim 10^{-3}$, for standard opacity choices). However, once the nominal mass-loss prescription is included, when the cavity exceeds approximately $\sim 30$~AU very rapid, run-away dust clearing occurs making the cavity optically thin when the cavity radius is about $\sim 40$~AU approximately $5\times10^5$~years after gas accretion ceases.   

\begin{figure}
\centering
\includegraphics[width=\columnwidth]{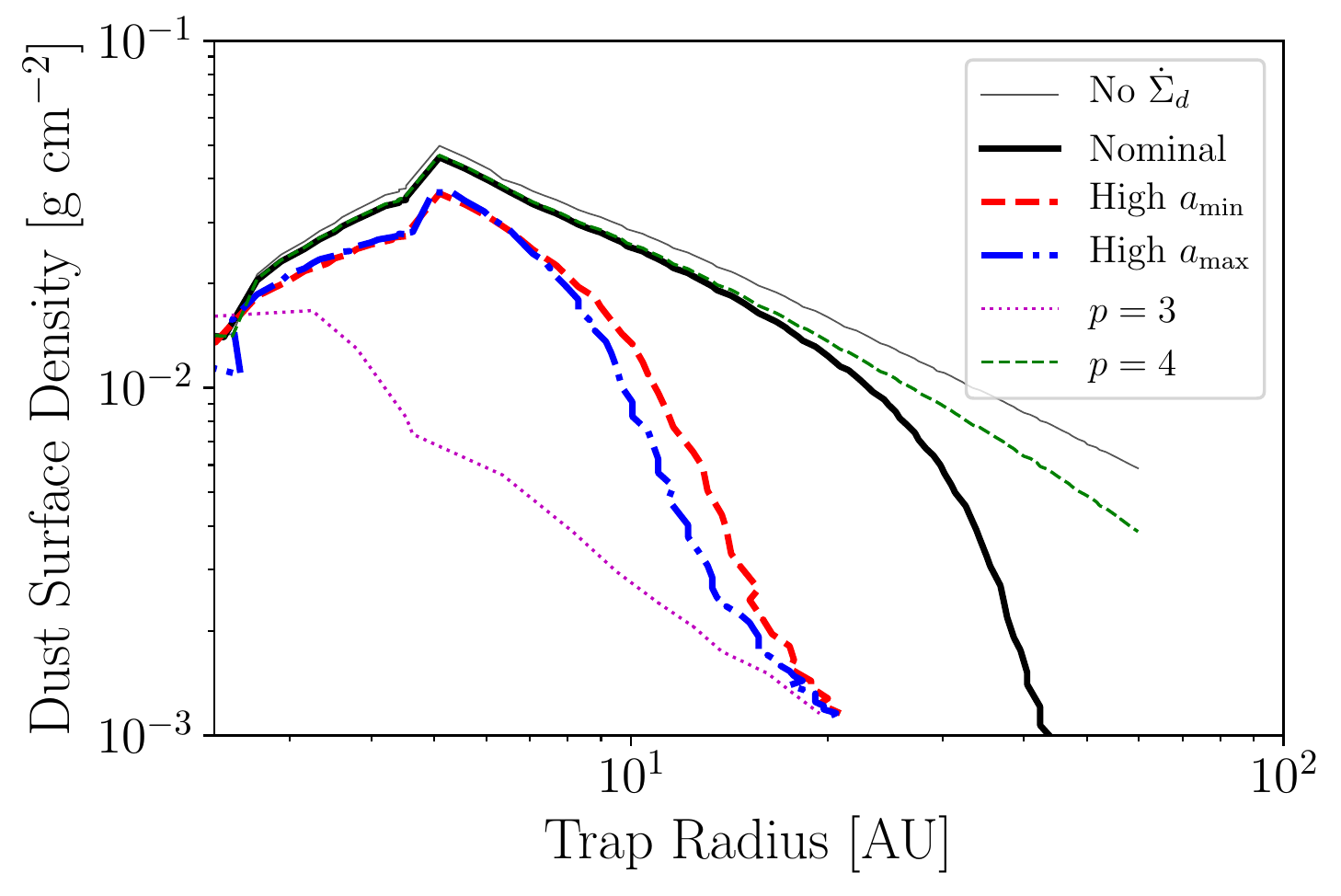}
\caption{The evolution of the dust surface density in the pressure trap as a function of radius, for the case of no dust-mass loss, or nominal model and models where the dust parameters are varied.}\label{fig:evolve_vary}
\end{figure}

We can see the dependence of the dust parameters explored previously in Figure~\ref{fig:evolve_vary}, where we show the dust surface density in the pressure trap as a function of pressure trap radius. 
This Figure shows that in all cases expect that with a bottom heavy dust size distribution ($p=4$) that rapid clearing of the pressure trap can proceed before the dust trap radii becomes large. Of particular note are the cases where we modify the dust populations away from the nominal case, by increasing $a_{\rm min}$ or $a_{\rm max}$ or by giving the dust size population a top-heavy distribution. As hinted at by Figure~\ref{fig:vary_parameters}, we end up with very rapid clearing. The timescale for clearing is not $\ll 10^5$~years as 
indicated by Figure~\ref{fig:vary_parameters}. Radial drift resupplies lost dust-mass back into the trap; however, in these cases the dust trap becomes radially optically thin by the time the trap radius has reached $\sim 10$~AU (note that since these models have lower opacity overall than the nominal case the trap becomes optically thin at surface densities around $3\times10^{-3}$~g~cm$^{-2}$ rather than the $\sim 10^{-3}$~g~cm$^{-2}$ value for the nominal model). 

Finally, as the models of \citet{Owen2011b} identified those young stars with lower than average photoevaporation rates (due to low stellar X-ray luminosity) as the most prevalent cause of long-lived relic discs, we run one final model using the nominal dust parameters for a disc with an X-ray luminosity of $3\times10^{29}$~erg~s$^{-1}$.    

\begin{figure}
\centering
\includegraphics[width=\columnwidth]{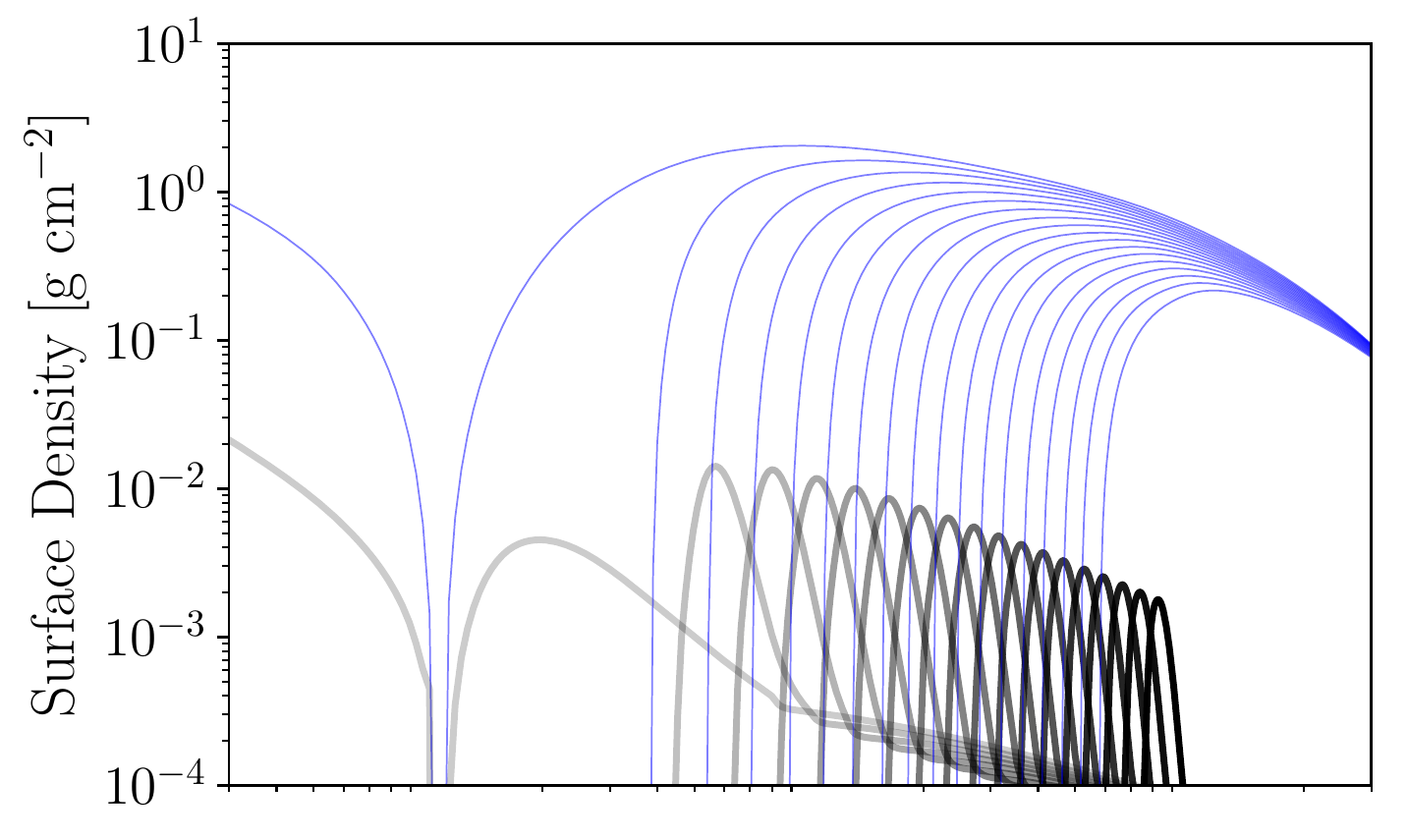}
\includegraphics[width=\columnwidth]{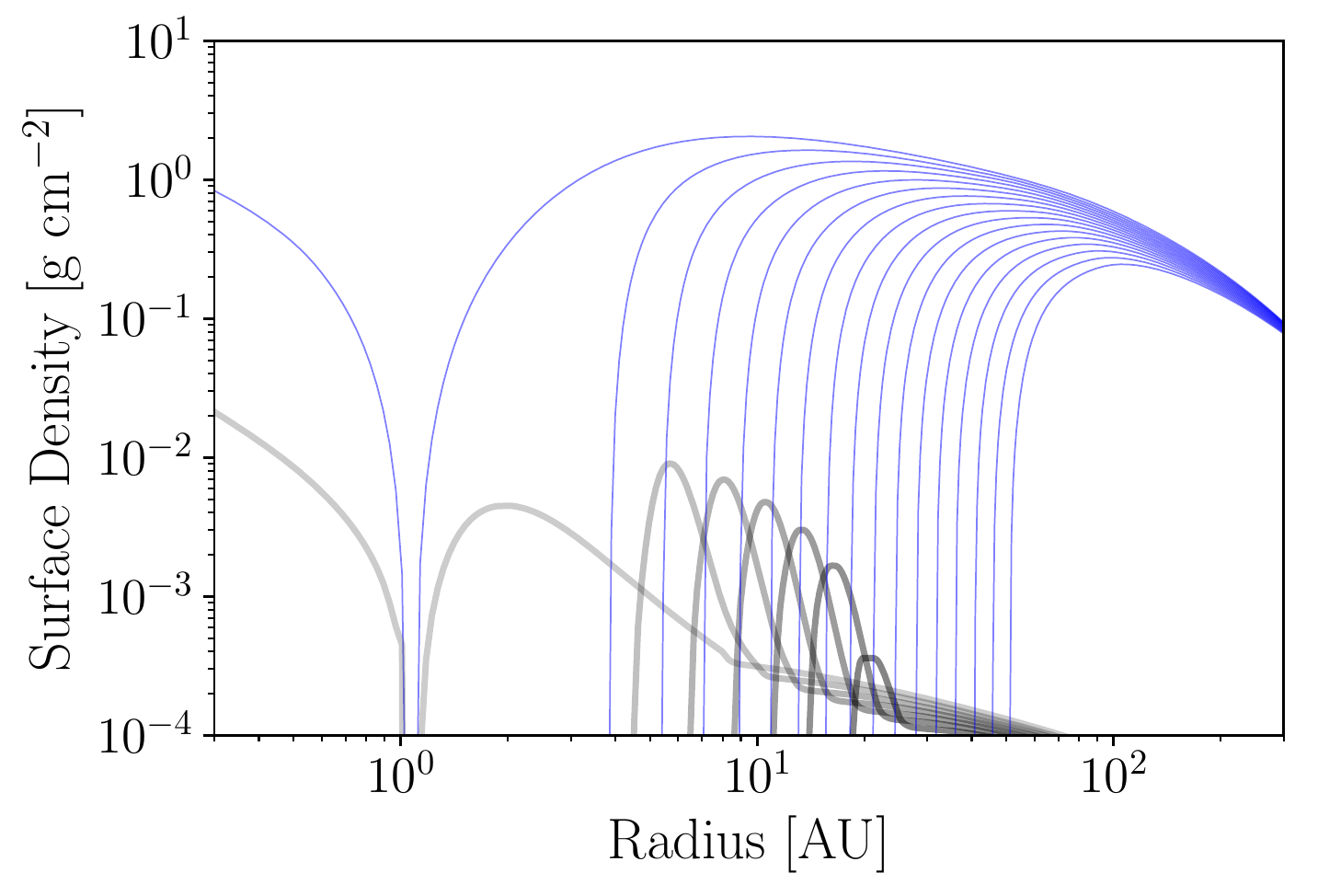}
\caption{The evolution of the dust (black, thick) and gas (blue, thin) surface densities in a photoevaporating disc, with a low X-ray luminosity of $3\times10^{29}$~erg~s$^{-1}$. The top panel shows a model with no radiation pressure clear-out. The bottom panel is the same model, but including dust mass-loss for the ``nominal'' dust parameters. The model is initially shown just after gap opening (after $\sim 8.27$~Myr of evolution) and then every 0.2Myr. {\bc The darkness of the dust lines increases with time.} }\label{fig:s_evolve_low_lx}
\end{figure}

For this low X-ray luminosity case, the dust cavity is cleared when the cavity radius reaches $\sim 20$~AU, even with the nominal dust parameters. The reason for the relatively more rapid clearing is that the dust surface densities initially present in the trap are lower than our standard case. This lower surface density is because lower X-ray luminosities result in lower photoevaporation rates which mean more disc material has accreted onto the star by the time photoevaporation has cleared the inner disc. Therefore, radiation pressure driven dust clearing can proceed efficiently around low X-ray stars as well, modifications to the dust population discussed above like higher minimum grain sizes or larger fragmentation velocities are only going to make clearing even more efficient. 
\subsection{Observational signatures}
\begin{figure}
\centering
\includegraphics[width=\columnwidth]{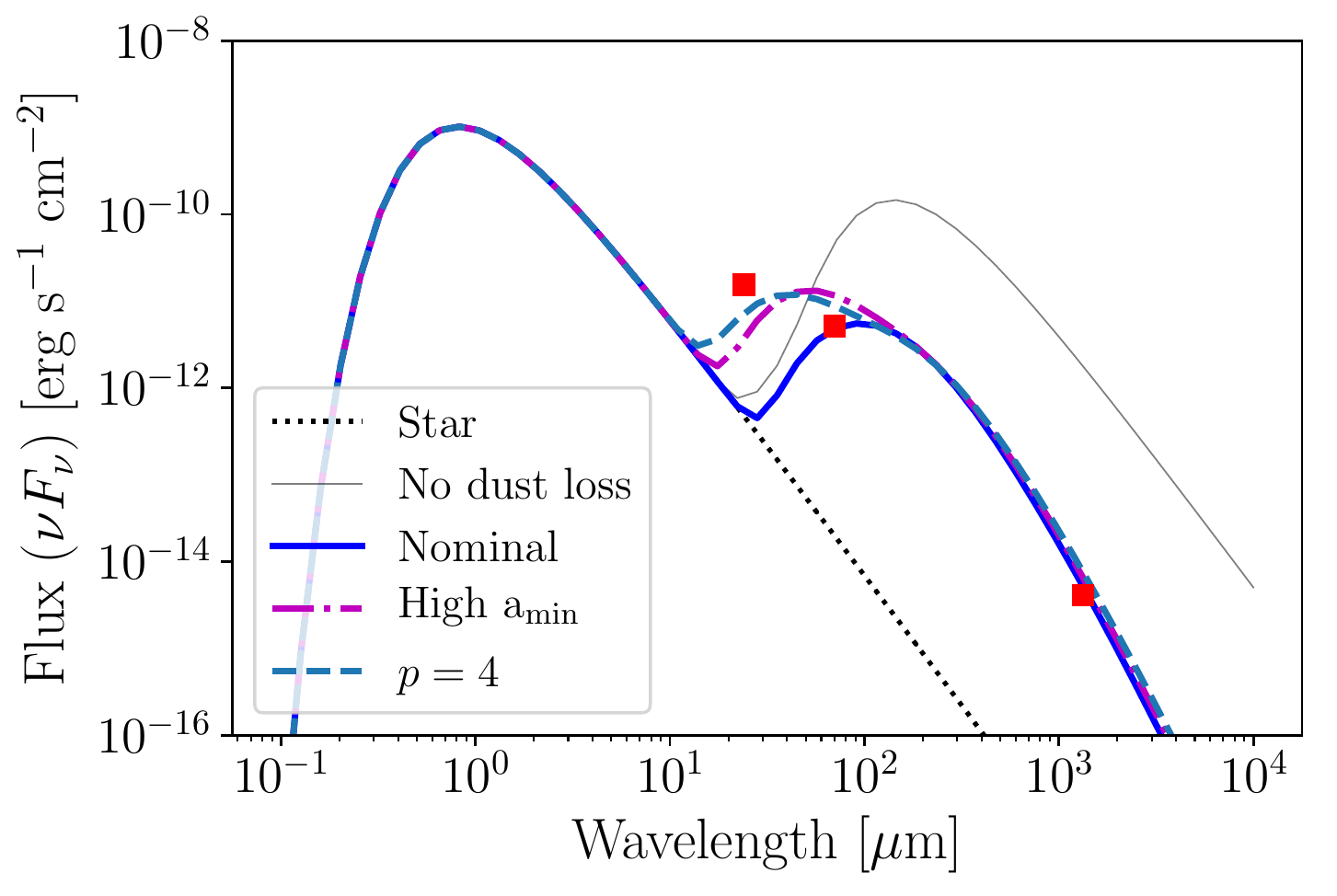}
\caption{{\bc The spectral energy distributions of the young star and disc for disc evolution scenarios without dust-mass loss due to radiation pressure (thin line) and  for various models including dust mass-loss due to radiation pressure (thick lines). The stellar SED is shown as the dotted line. The square points show the {\it Spitzer} MIPS and ALMA fluxes of the typical non-accreting transition disc SZ112 with fluxes taken from \citep{Romero2012,Hardy2015}.}}\label{fig:SEDs}
\end{figure}

This work was motivated by the fact that the standard photoevaporation model (e.g. top panels of Figures~\ref{fig:s_evolve} \& \ref{fig:s_evolve_low_lx}) predicts transition discs which are not accreting but have radially optically thick walls near the cavity edge which give rise to large MIR/FIR excesses above the photosphere \citep[e.g.][]{Clarke2001,Alexander2006b,Owen2011b}. Transition discs with these characteristics are not observed \citep{Owen2012,Cieza2013,Hardy2015,Owen2016}. Therefore, in this section, we compute spectral energy distributions of our discs on the point at which the pressure trap has become radially optically thin.  Since the goal of this section is to compute representative SEDs we do not use a full numerical radiative transfer approach, rather we estimate the disc's radial temperature profile and then calculate the spectral energy distribution as:
\begin{equation}
\lambda F_{\lambda}= \frac{\lambda}{d^2}\int_0^\infty2\pi R B_\lambda(T(R))\left[1-\exp\left(-\tau_\lambda\right)\right]{\rm d}R
\end{equation}
where $d$ is the distance to the source, which we set to 150~pc (corresponding to a typical distance to a young star in the Gould Belt) and $B_\lambda$ is the Planck function. To estimate the radial temperature profile we smoothly match an optically thin temperature profile to an optically thick temperature profile \citep[e.g.][]{Owen2014}, adopting the black-body temperature for optically thin radiation and the optically thick temperature profile used by \citet{Owen2011b}. In this case, our temperature profile becomes:
\begin{eqnarray}
T(R) &=& T_{\rm BB}\left(\frac{R}{1~{\rm AU}}\right)^{-1/2}\exp\left(-\tau_*\right)\nonumber\\
&& +100\,{\rm K}\left(\frac{R}{1~{\rm AU}}\right)^{-1/2}\left[1-\exp\left(-\tau_*\right)\right]
\end{eqnarray}
where $T_{\rm BB}$ is the black-body temperature, and $\tau_*$ is the mid-plane optical depth to the stellar irradiation. If there is a radially optical thick wall that occurs at a gap edge, we include an additional blackbody component for this wall at its local temperature.   The opacities are calculated as a function of radii using our maximum grain size determined from the dust evolution algorithm and a power-law dust size distribution with $p=3.5$ with a minimum grain size of 0.1$\mu$m (irrelevant of the actual dust parameters chosen for the trap). The absorption efficiencies are calculated identically to those described in Section~\ref{sec:Numerical}. The resultant SEDs for the nominal model, high $a_{\rm min}$ (the high $a_{\rm max}$ SED looks similar to this model) and model with dust size distribution parameter $p=3$. They are plotted in Figure~\ref{fig:SEDs} at the point when the dust in the pressure trap becomes radially optically thin. This point occurs at a dust trap radius of $\sim 40$~AU and time of 4.25~Myr for the nominal model; $\sim 15$~AU and 3.76~Myr for the high $a_{\rm min}$ case and $\sim 8$~AU and 3.59~Myr for the $p=3$ model. {\bc Additionally, we also include an SED for a model which does not include any dust mass-loss due to radiation pressure (i.e. the original photoevaporation model of \citealt{Owen2011b}), this is shown when the hole radius is at 10~AU and is clearly inconsistent with the observations. This disc has a radially optically thick dust wall at the gap edge, and the disc's SED is dominated by MIR emission from this wall which is reprocessing a large fraction of the star's luminosity \citep[e.g.][]{Ercolano2015}. }
These SEDs compare favourably to the photometry reported by \citet{Romero2012,Cieza2013} and \citet{Hardy2015} for non-accreting stars that also showed an IR excess, who reported MIR/FIR fluxes in the range $\sim 10^{-12}$ erg~s$^{-1}$~cm$^{-2}$ and {\it ALMA} 1.3mm fluxes of order $10^{-15}$ erg~s~cm$^{-2}$ for a sample of nearby young stars at a distance of 150~pc. {\bc In the Figure, we explicitly show {\it Spitzer} MIPS and ALMA photometry for the non-accreting transition disc SZ112 which is consistent with the models which include dust-mass loss due to radiation pressure.} Therefore, our model produces SEDs consistent with what has been previously classified as young debris discs. This classification is not surprising as young stars identified as hosting young debris discs are done so because they show an optically thin SED.   
\section{Discussion}\label{sec:discuss}
We have studied the impact of radiation-pressure on clearing dust traps created by photoevaporation. Small dust particles are removed from the pressure trap in the upper layers of the disc, which are then replaced by fragmentation of larger particles in the mid-plane. This coupling of radiation pressure removal to fragmentation means that radiation pressure is effectively able to remove dust particles of all sizes. This coupling will cease to be effective when the collision timescale for the largest grains becomes longer than the vertical diffusion timescale. The collision timescale for largest grains is $\sim \Omega_K^{-1}(\Sigma_g/\Sigma_d)$, while the vertical diffusion timescale is $\sim \Omega_K^{-1}(Z_{\rm phot}/H)^2/\alpha$. Therefore, provided the dust-to-gas ratio is $\gtrsim 10^{-3}$ which is generally true for optically thick dust traps the collision frequency of the largest grains is high enough that we do not need to worry about the resupply of smaller grains to photospheric heights (typically 2-3 $H$, see Figure~\ref{fig:1d_2d_compare}). Therefore, the clearing of discs is either limited by the diffusion timescale of small particles from the mid-plane to above the photosphere or by the clearing time-scale of small particles from above the photosphere themselves. 

Our calculation indicates that unless the parameters of the system are such that the opacity of the dust trap to incoming stellar photons is higher than we have estimated then radiation pressure clearing is an essential ingredient of disc dispersal, potentially solving the issue of long-lived ``relic'' discs. Clarity on this issue of whether this model does indeed satisfy all existing observational constraints must, unfortunately, wait until more sophisticated calculations are completed (see Section~\ref{sec:future}). However, we can sketch out some of the basic concepts below. Once photoevaporation opens a gap and causes a pressure trap, dust will rapidly grow and become trapped in said pressure trap. Unlike the primordial disc case \citep{Takeuchi2003}, once the pressure trap forms, the photosphere will no-longer intercept the disc again at large radius and radiation pressure begins clearing small dust particles from the disc. This process leads to a run-away reduction in the surface density in the disc until radial drift of new dust-particles can refill the dust-trap, whence the dust-to-gas ratio is low, and the disc's mid-plane is radially optically thin to a large radius. Hence the final stages of the disc's lifetime will be characterised by a dust-depleted, gas-rich disc which contains somewhere in the region of $\sim 0.1$~M$_{\rm jup}$ of gas. Photoevaporation of this disc will then continue to erode the disc to a large radius; although the thermodynamics of this new dust-depleted disc needs to be studied to determine if thermal sweeping is relevant. Such a disc would be observationally classified as a gas-rich, cold debris disc; where the origin of the dust is primordial, rather than secondary as often assumed. 

\subsection{Millimetre-bright transition discs}\label{sec:mmbright}
We must comment on how this process may operate in mm-bright transition discs. Millimetre-bright transition discs are those discs which are also thought to contain a pressure trap that is created through some unknown mechanism, potentially planet-disc interactions \citep[e.g.][]{Espaillat2014,Owen2016}. The observed dust surface densities in these pressure traps are more of the order of 1~g~cm$^{-2}$ \citep[e.g.][]{Andrews2011,Owen2017}. Since these objects are observed to be long-lived radiation pressure clear-out cannot be fast in these objects. The very high-surface densities imply the photosphere sits at a large height in the disc, indicating the dust mass-fraction above the photosphere is small, and hence the dust clearing timescales are long. A simple 1D calculation for parameters typical of mm-bright transition discs indicates they are stable on Myr timescales. 

These long depletion times do not mean that radiation pressure clear-out is not interesting, in-fact one of the mysteries of mm-bright transition discs is the lack of small grains close to their star. We speculate if radiation pressure could remove small dust particles from the surface layers of their dust-traps before they can drift into the inner disc then this could provide a solution to this outstanding problem. 

\subsection{Future direction} \label{sec:future}
While our results are promising and indicate that radiation pressure driven mass-loss is vital in disc dispersal, it is not entirely clear what the exact dust mass-loss rates are. This uncertainty is because the mass-loss rates are sensitive to parameters we have merely assumed rather than calculated self-consistently. The minimum and maximum grain sizes, as well as the dust-particle size distribution, are assumed in all our above calculations. To model radiation pressure clear-out all these parameters need to be calculated explicitly because the mass-loss essentially depends on the dust mass-fraction above the photosphere, but the photospheric position depends on the 2D particle size distribution. Therefore, to accurately calculate the clearing timescales and observable properties we need to incorporate a dust evolution algorithm and thermal structure solver into our 2D code presented in this paper.  
\section{Summary}
We have introduced a new disc dispersal mechanism that takes place in the final stages of a disc's lifetime. Our new mechanism focuses on the removal of dust from the pressure trap that is created when photoevaporation opens a cavity in an evolved protoplanetary disc. We find that radiation pressure can efficiently remove small particles from the surface layers of the disc, in the vicinity of the pressure trap; these small particles are then replaced by the collisional fragmentation of larger particles in the mid-plane. With this new dust mass-loss process the photoevaporation model (in particular the X-ray photoevaporation model) does not suffer from a severe ``relic'' disc problem, as optically thick discs with hole sizes of $\gtrsim 50$~AU are never created.

The photoevaporation created dust traps are depleted on timescales ranging from $\lesssim 10^5$~years while the cavity radius is $\lesssim 10$~AU to a few $10^5$~years by the point the cavity radius has receded to $\sim 40$~AU. Clearing of dust from the pressure traps proceeds until the point where radial-drift from the outer regions of the disc balances radiation pressure driven loss, but by this point, the disc's mid-plane is radially optically thin to a large radius. The clearing time is sensitive to the opacity structure, and therefore particle size distribution in the vicinity of the pressure trap, where lower opacities results in faster clearing timescale. 

Ultimately, the combination of photoevaporation and radiation pressure driven mass-loss results in a disc which observationally appears as a young gas-rich debris disc. Indeed many of the previously identified gas-rich debris discs could indeed be discs in this stage of their clearing, where the dust is primordial in origin.  We hypothesise a sensitive CO survey of young Weak T Tauri stars may find a significant fraction of them that still host $\sim 0.1$~M$_{\rm jup}$ gas reservoirs at large radii.

\section*{Acknowledgements}

We thank the referee for a constructive report which improved the manuscript. We are grateful to Richard Booth, Cathie Clarke, Ruth Murray-Clay and Giovanni Rosotti for interesting discussions. JEO is supported by a Royal Society University Research Fellowship.







\bsp	
\label{lastpage}
\end{document}